\documentclass[journal]{IEEEtran}
\usepackage{amsmath}
\usepackage{amsfonts,amssymb}
\usepackage{mathrsfs}
\usepackage{bm}
\usepackage{indentfirst}
\usepackage{graphicx}
\usepackage{epstopdf}
\usepackage[font=small, labelsep=period]{caption}
\usepackage[ruled,linesnumbered]{algorithm2e}
\usepackage{color}
\usepackage{booktabs}
\usepackage{threeparttable}
\usepackage{url}
\usepackage{cite}
\usepackage{subfigure}
\usepackage{array}
\usepackage{multirow}
\usepackage{makecell}

\begin{document}

\title{IMRecoNet: Learn to Detect in Index  Modulation Aided MIMO Systems with Complex Valued Neural Networks}

\author{Chenwu~Zhang,
		Hancheng~Lu,~\IEEEmembership{Senior Member,~IEEE,} 
		and Jinxue~Liu
		\IEEEcompsocitemizethanks{
			\IEEEcompsocthanksitem This work was supported in part by the National Science Foundation of China (No.61771445, No.61631017, No.91538203). 
			
			Chenwu~Zhang, Hancheng Lu and Jinxue Liu are with School of Information Science and Technology, University of Science and Technology of China, Hefei 230027, China (email: cwzhang@mail.ustc.edu.cn; hclu@ustc.edu.cn; jxliu18@mail.ustc.edu.cn).
		}
	}

\maketitle

\begin{abstract}
Index modulation (IM) reduces the power consumption and hardware cost of the multiple-input multiple-output (MIMO) system by activating part of the antennas for data transmission. However, IM significantly increases the complexity of the receiver and needs accurate channel estimation to guarantee its performance. To tackle these challenges, in this paper, we design a deep learning (DL) based detector for the IM aided MIMO (IM-MIMO) systems. We first formulate the detection process as a sparse reconstruction problem by utilizing the inherent attributes of IM. Then, based on greedy strategy, we design a DL based detector, called IMRecoNet, to realize this sparse reconstruction process. Different from the general neural networks, we introduce complex value operations to adapt the complex signals in communication systems. To the best of our knowledge, this is the first attempt that introduce complex valued neural network to the design of detector for the IM-MIMO systems. Finally, to verify the adaptability and robustness of the proposed detector, simulations are carried out with consideration of inaccurate channel state information (CSI) and correlated MIMO channels. The simulation results demonstrate that the proposed detector outperforms existing algorithms in terms of antenna recognition accuracy and bit error rate under various scenarios. \\
\end{abstract}

\begin{IEEEkeywords}
Index modulation, multiple-input multiple-output, signal detection, sparse reconstruction, complex-valued neural network.
\end{IEEEkeywords}

\section{Introduction}\label{introduction}
\IEEEPARstart{W}{ith} the rise of new mobile services, such as virtual reality, augmented reality and 4K video transmission, wireless air interface is facing tremendous pressure and challenges 
\cite{5G-Survey2,R1}. In order to further improve the throughput of the wireless network, the massive multiple-input multiple-output (MIMO) \cite{massiveMIMO1,massiveMIMO0} \textcolor{blue}{as well as  millimeter wave (mmWave) technology\cite{mmwavepower} have become}  the key technologies in the fifth generation (5G) wireless communication systems, which has attracted extensive and enthusiastic \textcolor{blue}{research} in recent years. It is undeniable that the massive MIMO does bring significant throughput and reliability improvements by operating the large-scale transmit antenna (TA) array at the base station in diversity or multiplexing mode. However, \textcolor{blue}{energy consumption in MIMO and mmWave technology have become a big concern in 5G wireless communication\cite{mmwavepower}.} As the number of TA grows, the huge power consumption \cite{massiveMIMOPower}  and expensive hardware cost caused by radio frequency (RF) chains bound to each TA also appear, which make the actual deployment of massive MIMO face severe challenges \textcolor{blue}{especially when the  mmWave communication\cite{mmwavepower}} is used. Hence, how to balance the performance of wireless communication systems with the power consumption and hardware cost is still an issue worthy of attention when the massive MIMO \textcolor{blue}{and mmWave communication technology are}  practically applied to current wireless communication systems.

As an assistive technology that can reduce the power consumption and hardware cost, index modulation (IM) \cite{IM-Mag} has attracted extensive research in recent years. IM is a general term for a series of innovative digital modulation technologies, which conveys information bits not only on the modulated symbols but also on the on-off status of some resource blocks (such as antennas, subcarriers and time slots) \cite{JSAC-SM,OFDM-IM,SM-Space-Time}. In this paper, we focus on the IM with MIMO, i.e., only part of the TAs are seletected to be active for data transmission based on the part of the information bits. Due to this unique property, it is obvious that IM can bring the  \textcolor{blue}{remarkable reduction of power consumption and hardware cost of transmitter in 5G communication}. Although IM is excellent in power consumption and hardware cost, the significant increase in receiver complexity caused by it cannot be ignored. In the IM aided MIMO (IM-MIMO) systems, not only the transmitted symbol but also the activation status of each TA need to be detected at the receiver, which greatly increases the complexity compared with the traditional MIMO detectors. More specifically, the increase in receiver complexity mainly comes from the detection of antenna activation status. Once the activation status of the TAs is detected, some linear MIMO detectors, such as least square (LS) and linear minimum mean square error (LMMSE), can be used to quickly detect the signal. Therefore, the detection of the activation status of TAs directly determines the complexity of the entire receiver.

For the detection in the IM-MIMO systems, several algorithms have been proposed over the past decades. \textcolor{blue}{In \cite{SM-MIMO}, Yang \emph{et al.} applied the deep learning technique to the signal detection of SM-MIMO system and problems were solved by data-driven prediction approaches.}
Jeganathan \emph{et al.} proposed an exhaustive search based algorithm \cite{SM-ML}, which has the optimal performance but with high complexity. To further reduce the complexity, several algorithms based on convex optimization \cite{SVBL} or probability model \cite{MIMO-OFDM-IM} have been investigated. 
However, performance, convergence and complexity issues hinder the practicality of these algorithms. In particular, imperfect channel state information (CSI) should be considered, for the reason that it is difficult to do perfect channel estimation when the number of TA becomes larger.  

This work addresses the detection of the IM-MIMO system from a different perspective. Instead of tackling the detection problem analytically, we leverage the recent advances in deep learning (DL) to design a data-driven algorithm to achieve a better performance with low complexity. In particular, the proposed approach establishes a connection between sparse reconstruction problem and minimizing a loss function in training a convolution neural network (CNN), rather than directly solve the detection problem of the IM-MIMO systems.

DL has gotten great success in computer vision, natural language processing and some other applications. Recent research has shown that DL has powerful potential in solving difficult wireless communication problems, such as channel estimation \cite{RoemNet}, signal detection \cite{IMNet} and power control \cite{DDPG-PC}. However, current DL libraries, such as Tensorflow and PyTorch, only support real value operations, which does not match the processing of complex signals in communication systems. A question thus arises: if complex value operations are adopted to the design of CNN, can we achieve better performance than the real value based CNN with the same structure?

In this paper, we answer this question by proposing a DL based detector, called IMRecoNet, which has been embedded complex value operations, to solve the detection problem of the IM-MIMO systems efficiently and robustly. This work is a further advancement of our previous work \cite{IMNet} that introduces the complex value operations into the design of neural networks. The proposed IMRecoNet has two unique features: one is that it is insensitive to the CSI, which means it can still maintain excellent performance under imperfect CSI. And the other is that it supports complex value operations to adapt the complex signals in wireless communication systems. The main contributions of this paper are summarized as follows.

\begin{itemize}
	\item By utilizing the unique attribute of IM, we model the transmission process of the IM-MIMO system as a compressed sensing (CS) process. Correspondingly, the detection process at the receiver can be modeled as a sparse reconstruction problem, i.e., the inverse process of CS. Then, we divide this sparse reconstruction problem into two sub-problems. One is the support\footnote{Set of indices corresponding to nonzero elements in vector $ \mathbf{s} $ is called the support $ \mathbf{\Omega}_\mathbf{s} $ of $ \mathbf{s} $. If $ \mathbf{s} = \left[ 0,2,0,0,3 \right] $, then $ \mathbf{\Omega}_\mathbf{s} = supp_{\textcolor{blue}{1}}(\mathbf{s}) = \left[ 1,4 \right] $. For matrix, $ \mathbf{X} = \left[ \mathbf{x}_1, \mathbf{x}_2, \cdots, \mathbf{x}_n \right] $, its support is $ \mathbf{\Omega}_\mathbf{X} = supp_{\textcolor{blue}{1}}(\mathbf{x}_1)\cup supp_{\textcolor{blue}{1}}(\mathbf{x}_2) \cup \cdots \cup supp_{\textcolor{blue}{1}}(\mathbf{x}_n) $.} identification (i.e., the detection of active TAs), and the other one is sparse estimation (i.e., the detection of transmitted symbols on the active TAs).
	\item To solve the sparse reconstruction problem (i.e., support identification and sparse estimation) with low complexity and good robustness, we design a DL based detector, called IMRecoNet, which combines two CNNs with linear MIMO detector (i.e., LS). IMRecoNet can directly predict the activation probability of each TA without iteration, which guarantees its low complexity. In addition, IMRecoNet does not need CSI as its input, which means it can perform well with inaccurate channel estimation, even without channel estimation. This feature guarantees its robustness in various scenarios. With the aid of IMRecoNet, linear MIMO detector can directly detect the signal with low complexity.
	\item Considering that signal processing in wireless communication systems is based on complex numbers, complex-valued operations (i.e., complex convolution and complex activation) are adopted to the design of IMRecoNet, which can extract amplitude and phase characteristics of complex signal.
	\item In order to verify the adaptability and robustness of the proposed IMRecoNet, we carry out simulations with the consideration of inaccurate channel estimation at the receiver and the correlated MIMO channel. Simulation results demonstrate that the proposed IMRecoNet performs as well as traditional algorithms under perfect CSI, while the performance of the proposed IMRecoNet is far superior to traditional algorithms under imperfect CSI and correlated MIMO channel. Apart from this, the IMRecoNet that introduces complex-valued operations also has shown better performance than the real number network with the same structure.
\end{itemize}

The rest of this paper is organized as follows. Related work are discussed in Section \ref{related-work}. The system model and problem formulation are given in Section \ref{system-model-problem-formulation}. Section \ref{IMRecoNet} details the architecture of the proposed IMRecoNet and the implementation of adopting complex value operations to the design of IMRecoNet. The simulation results are provided in Section \ref{Performance-Evaluation}. Finally, conclusions are drawn in Section \ref{Conclusion}.

\emph{Notation:} Vectors and matrices are denoted by lowercase and uppercase bold letters, respectively. $ \left( \cdot \right)^T $, $ \left( \cdot \right)^H $, and $ \left( \cdot \right)^{-1} $ stand for transpose, Hermitian transpose and matrix inversion operations, respectively. 
$ \| \cdot \|_0 $ denotes the Zero norm,
$ \| \cdot \|_1 $ denotes the Taxicab norm,
$ \| \cdot \|_2 $ denotes the Euclidean norm, while $ \| \cdot \|_F $ is the Frobenius norm. $ \langle \mathbf{x}, \mathbf{y} \rangle $ denotes the inner product between vector $ \mathbf{x} $ and vector $ \mathbf{y} $. $ \mathcal{CN}(0,\sigma^2) $ represents the complex Gaussian distribution with zero mean and $ \sigma^2 $ variance. \textcolor{blue}{$C_{N}^{K}$} and $ \lfloor \cdot \rfloor $ denote the binomial coefficient and floor function, respectively. 
\textcolor{blue}{$ supp_{1}(\cdot) $ represents the support of vectors(i.e.  the indices of non-zero element) and $ supp_{2}(\cdot) $ represents the support of matrices(i.e. the detection of transmitted symbols on the active TAs).} $ \varnothing $ represents the empty set. $ \circledast $ is the convolution operator. $ \Re \left\{ \cdot \right\} $ and $ \Im \left\{ \cdot \right\} $ denote the real part and the imaginary part of a vector or matrix, respectively. $ Cov(\cdot) $ represents the covariance operation.

\section{Related Work}\label{related-work}
\subsection{Index Modulation Aided Wireless Communication Systems}
IM is an assistive technology aimed at reducing the power consumption and hardware cost, which has been widely applied to existing wireless communication systems in different ways. Research in \cite{OFDM-IM} and \cite{GSM} applied IM to MIMO systems and orthogonal frequency division multiplexing (OFDM) systems, respectively. To further improve the spectrum efficiency, literature \cite{MIMO-OFDM-IM} combined IM with OFDM (\textcolor{blue}{OFDM}-IM) and apply OFDM-IM to MIMO systems. Besides, applying IM across multiple domains (such as space, frequency and time domains) also attracts some attention and research \cite{Space-Time-TVT}\cite{Two-Dim-TVT} including integrating IM with non-orthogonal multiple access \cite{R3}\cite{R4}. IM reduces power consumption and hardware cost while significantly increasing the complexity of the receiver. It is worth noting that the traditional detectors in classical MIMO systems cannot be directly applied to the IM aided MIMO wireless communication systems. Since part of the information bits are conveyed on the indices of the active resource blocks, not only the modulated symbols but also the indices of the \textcolor{blue}{active} resource blocks need to be detected. Therefore, how to design an efficient detector for IM aided wireless communication systems is still a key issue.

Literature \cite{SM-ML} and \cite{OFDM-IM-ML} analyzed the maximum likelihood (ML) based detector, which has the optimal performance. Although the ML based detector has the optimal performance, it should not be ignored that its complexity increases exponentially with increase of the number of TAs and the signal constellation size. Attentive to the high computational complexity of the ML detection, researchers proposed other methods for the detection of the IM-MIMO systems. Mesleh \emph{et al.} proposed the matched filtering detection \cite{SM}. This method decouples the active antenna detection and the modulated symbol detection. Specifically, the active antennas are first estimated, and then, the modulated symbols are estimated. Literature \cite{SVBL} and \cite{SVBL2} proposed a detection scheme based on the angle between its corresponding channel vector and the received signal. In \cite{MIMO-OFDM-IM}, authors investigated two types of detection algorithms based on the sequential Monte Carlo theory, which achieve near-optimal error performance while maintaining low computational complexity.

The detection becomes more difficult if imperfect CSI is considered at the receiver. All the methods mentioned before assume that the receiver has known perfect CSI. However, perfect channel estimation is difficult and takes up more system resources. From a more practical perspective, imperfect CSI should be considered in the design of the detector for IM-MIMO systems.

\subsection{Deep Learning Meets Wireless Communication Systems}
In view of the great success of DL in various aspects, researchers began to apply DL to wireless communication systems in \cite{DL-5G}. From physical layer to MAC layer, many works have appeared.

Channel estimation is a typical case of successful application of DL in wireless communication systems. Authors in \cite{RoemNet} introduced DL to the channel estimation in OFDM systems. To tackle the challenge that traditional DL method is not reliable enough when the conditions of online deployment of the common neural networks are not consistent with the channel models used in the training stage, they proposed a novel meta learning based channel estimation. Considering the multiple channels as multiple tasks, a meta-learner which could learn the general characteristics of these tasks was trained off-line. When deploying the approach online, the meta-learner will be used to guide the adjustment of a new network by using a few pilots for fine-tuning. For exploiting the expert knowledge in wireless communications, Gao \emph{et al.} proposed a model-driven DL approach \cite{ComNet}. This model-driven approach can overcome the heavy demand of a huge amount of training data and use expert knowledge to facilitate further performance improvement. Besides, DL has the potential to improve performance in the areas of wireless communications, such as channel decoding \cite{ChannelDec-DL}, CSI feedback \cite{DL-CSI-Feedback}, precoding \cite{DL-Precoding} and end-to-end communication \cite{DL-End-to-end}.

Narrow down to the detection of wireless communication systems, authors in \cite{DL-MIMO-Det} investigated a model-driven DL detector for MIMO systems. In particular, this DL based detector is designed by unfolding an iterative algorithm and adding some trainable parameters. Compared with the data-driven DL based detector, this model-driven DL based detector can be rapidly trained with a much smaller dataset, while  the channel estimation taken into consideration. Similar to \cite{DL-MIMO-Det}, literature \cite{DeepMIMO} also proposed a model-driven DL based detector by unfolding the iteration of a projected gradient descent algorithm, which has comparable performance with approximate message passing based detector and is more robust to ill-conditioned channels. For the IM aided wireless communication systems, there are some research on the design of DL based detector. Authors in \cite{DL-IM-OFDM} proposed a CNN based detection framework for OFDM-IM systems, which achieved near ML detector performance with much lower complexity. Specifically, the received symbols are transformed to polar coordinates to help CNN detect the indices of active subcarriers and 2-dimensional convolution layers are utilized to fully exploit the inherent information in the received symbols. \textcolor{blue}{Besides, authors in \cite{ML-GSM} proposed a deep learning based detection scheme  for the visible light communication system using generalized spatial modulation, where a deep neural network consisting of several neural layers is applied to detect the received signals.}

There is no doubt that the research of DL in wireless communication system is currently enthusiastic. However, existing DL libraries, such TensorFlow and PyTorch, only support real value operations, while signals in wireless communication systems are often represented as complex values. When training the network, we have to split the complex number into two parts, i.e., the real part and the imaginary part. In recent years, \textcolor{blue}{research} on complex valued CNN has appeared. Authors in \cite{DeepComplexNN} proposed several key components in complex valued CNN, such as complex convolution, complex activation and complex batch normalization. The CNN embedded with complex value operation has shown great potential on audio-related tasks. Besides, literature \cite{Complex-MRI} and \cite{Complex-beamformer} applied complex valued CNN to magnetic resonance imaging domain and speech enhancement domain, respectively.

\begin{figure*}[htbp]
	\centering
	\includegraphics[width=0.6\linewidth]{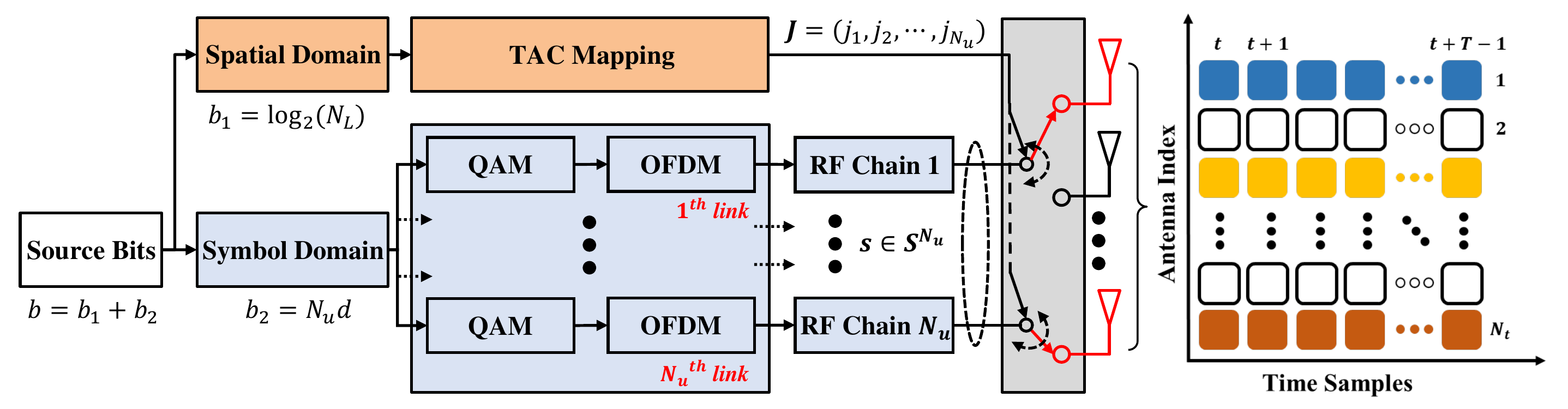}
	\captionsetup{name={Fig.},labelsep=period,singlelinecheck=off,font={small}}
	\caption{The block diagram of the transmitter of the IM-MIMO system, and the structure of transmitted signal in spatial domain and time domain. For the structure of transmitted signal, the solid colored squares represent the transmitted complex signal, while hollow colorless squares represent zero. And the $i$-th row represents the transmitted signal through the $i$-th TA in time domain, and the $j$-th column represents the transmitted signal vector for all TAs at the $j$-th transmit slot of a frame. Thus, the blank rows indicate that their corresponding TAs are off and do not send anything during a frame.}
	\label{fig-system-model}
\end{figure*}

\section{System Model and Problem Formulation}\label{system-model-problem-formulation}
In this section, the system model of IM-MIMO is first given, where IM is applied to the spatial domain. Then, the joint sparsity of the transmitted signal in IM-MIMO systems, which is introduced by IM, is analyzed. And the detection process at the receiver is formulated as a sparsity reconstruction problem. 

\subsection{System Model}\label{system-model}
We consider an IM-MIMO system that equipped with $ N_{t} $ transmit antennas (TAs), of which block diagram is depicted in Fig.\ref{fig-system-model}. Different from the traditional MIMO systems, only $ N_u \ (0 < N_u < N_t) $ TAs are active for data transmission in IM-MIMO systems, and $ N_u $ active TAs are selected by part of the information bits. Obviously, there exist $C_{N_t}^{N_u}$ TA combinations (TAC) for selecting $N_u$ TAs from $N_t$ TAs. In order to facilitate data mapping, only $N_L = 2^{\lfloor \log_2(C_{N_t}^{N_u}) \rfloor}$ TACs are legal and generated according to \cite{IM-Mag} based on $ N_t $ and $ N_u $ in advance. We denote the $N_L$ legal TACs as $\mathbb{J} = \left\{J_1,J_2,\cdots,J_{N_L}\right\}$.

As shown in Fig.\ref{fig-system-model}, each IM-MIMO frame is comprised of a total number of $ b=b_1+b_2 $ bits. The first $b_1=\log_2(N_L)$ bits are called spatial bits and used to select a TAC $J = \left\{j_1, j_2, \cdots, j_{N_u}\right\}$ from the TAC set $\mathbb{J}$, where $j_1, j_2, \cdots, j_{N_u} \in \left\{1,2,\cdots,N_t\right\}$ are the index of active TAs. Baesed on TAC $J$, antenna activation pattern (AAP) $\mathbf{g} \triangleq \left[g_1, g_2, \cdots, g_{N_t}\right]$ is defined as
\begin{equation}
	\label{form:aap}
	g_{\color{blue}k}=\begin{cases}
		0, & {\color{blue}k} \notin J \\
		1, & {\color{blue}k} \in J
	\end{cases}
\end{equation}
which $``1"$ means the corresponding TA is active and $``0"$ means the corresponding TA is inactive. For example, when $ N_t = 4 $ and $ N_u = \textcolor{blue}{2} $, the mapping relationship between the input binary bits, the TAC and the AAP is shown in Table. \ref{Table-AAP}.

\begin{table}[htbp]
	\centering\small
	\caption{The mapping relationship between the input binary bits, the TAC and the AAP with $N_t = 4, N_u = \textcolor{blue}{2}$.}
	\label{Table-AAP}
	\begin{tabular}{cccc}
		\hline\hline
		$\mathbf{b} = [b_0, b_1]$ & TAC Index   & TAC & AAP  \\
		\hline
		$[0\ 0]$ & 0 & $\left\{1,3\right\}$ & $[1,0,1,0]$  \\
		$[0\ 1]$ & 1 & $\left\{1,4\right\}$ & $[1,0,0,1]$  \\
		$[1\ 0]$ & 2 & $\left\{2,4\right\}$ & $[0,1,0,1]$  \\
		$[1\ 1]$ & 3 & $\left\{2,3\right\}$ & $[0,1,1,0]$  \\
		\hline\hline
	\end{tabular}
\end{table} 

Since only $N_u$ TAs are activated for actual information transmission, the IM-MIMO system only needs to be configured with $N_u$ signal processing links. After selecting the TAC, the left $ b_2 = N_ud $ bits are equally split to $ N_u $ blocks and processed on $ N_u $ signal links, which means each link processes $ d $ bits. After quadrature amplitude modulation (QAM) with constellation $ \mathcal{S} \left( |\mathcal{S}| = M \right)  $ and OFDM, the signals on $ N_u $ links are transmitted through $ N_u $ active TAs, respectively. We assume that an IM-MIMO frame is transmitted through $ T $ time slots. After passing through the MIMO channel, the received signal vector at the $j$-th $ \left( j \in \left[ 1, T \right]  \right)  $ time slot of the $i$-th IM-MIMO frame can be represented as
\begin{equation}
	\label{formula-transmission}
	\mathbf{y}_{j, i}=\mathbf{H}_{j,i}\left( J_i \right) \mathbf{s}_{j,i}+\mathbf{n}_{j,i},
\end{equation}
where $ \mathbf{H}_{j,i} = \left[ \mathbf{h}_1, \mathbf{h}_2, \cdots, \mathbf{h}_{N_t} \right]  \in \mathbb{C}^{N_r \times N_t} $ and $ \mathbf{s}_{j,i}\in\mathcal{S}^{N_u\times1} $ denote the MIMO channel matrix and actually transmitted signal vector at the $j$-th time slot of the $i$-th IM-MIMO frame, repectively. $ \mathbf{n}_{j,i} \in \mathbb{C}^{N_r\times1} $ represents the additive white Gaussian noise (AWGN) vector with zero mean and unit variance at the $ j $-th time slot in the $i$-th IM-MIMO frame. Particularly, $ J_i $ is the selected TAC for the $i$-th IM-MIMO frame, and $\mathbf{H}_{j,i}\left( J_i \right)=[\mathbf{h}_{j_1},\mathbf{h}_{j_2},\cdots,\mathbf{h}_{j_{N_u}}]$ consists of the columns of $\mathbf{H}_{j,i}$ indexed by $J_i$.

Since the information bits are conveyed on the TAC $J_i$ and actually transmitted signal vector $\mathbf{s}_{j,i}$, both $J_i$ and $\mathbf{s}_{j,i}$ should be detected at the receiver. Based on Eq.(\ref{formula-transmission}), the maximum likelihood detector (MLD) \cite{SM-ML} can be formulated as
\begin{equation}
\label{formula-MLD}
<\hat{J}_i, \mathbf{\hat{s}}_{j, i} > = \mathop{\arg\min}_{J_i \in \mathbb{J}, \mathbf{s}_{j,i} \in \mathcal{S}^{N_u}}{\|\mathbf{y}_{j,i}-\mathbf{H}_{j,i}\left( J_i \right)\mathbf{s}_{j,i}\|}_2^2.
\end{equation}
It should be noted that the MLD jointly detects $J_i$ and $\mathbf{s}_{j,i}$, which entails an exhaustive search space. Therefore, there will be a prohibitive complexity with large $N_t$, $N_u$ and $M$.

\subsection{Problem Formulation}\label{problem-formulation}
Let define the following mapping relationship
\begin{equation}
	\label{formula-mapping}
	\mathbf{s}_{j,i}\in\mathbb{C}^{N_u \times 1} \xrightarrow{J_i} \mathbf{x}_{j,i}\in\mathbb{C}^{N_t \times 1},
\end{equation}
which means that inserting $ N_t - N_u $ zeros to $ \mathbf{s}_{j,i} $ according to the TAC $ J_i $. By doing this mapping, all the TAs transmit information according to the following rules: the $ N_u $ active TAs transmit the modulated symbols generated by $ N_u $ signal processing links, while $ N_t - N_u $ inactive TAs transmit zeros. 

Based on the mapping relationship expressed as Eq.(\ref{formula-mapping}), the structure of transmitted signal in an IM-MIMO frame can be depicted as the right of Fig.\ref{fig-system-model}. Due to the fact that there are only $ N_u $ TAs are activated during the data transmission of an IM-MIMO frame, it is clear that there are only $ N_u $ non-zero element in each transmitted signal vector (i.e., $ \mathbf{x}_{j,i}, j \in \left[ t, t + T - 1 \right] $), which has been shown in Fig.\ref{fig-system-model}. When $ N_u $ is far less than $ N_t $ (i.e., $ N_u \ll N_t $), $ \mathbf{x}_{j,i} $ can be considered as a sparse vector and the sparsity is $ N_u $. For each IM-MIMO frame, the TAC is only selected once, which means the position of $ N_u $ non-zero elements in each transmitted signal vector of an IM-MIMO frame are same. Based on the mapping relationship in Eq.(\ref{formula-mapping}) and the structure of the transmitter shown in Fig.\ref{fig-system-model}, we have
\begin{equation}
	supp_{\textcolor{blue}{1}}(\mathbf{x}_{j,i}) = supp_{\textcolor{blue}{1}}(\mathbf{g}_i) = J_i, \forall j \in \left[1, T\right] \label{formula-supp1},
\end{equation}
\begin{equation}
	{\|supp_{\textcolor{blue}{1}}(\mathbf{x}_{j,i}) \|}_0 = {\| \mathbf{g}_i \|}_0 = N_u, \forall j \in \left[1, T\right] \label{formula-supp2},
\end{equation}
where $ supp_{\textcolor{blue}{1}}(\cdot) $ denotes the support (i.e., the indices of non-zero element) of a vector. Eq.(\ref{formula-supp1}) means all the transmitted signal vectors in the same IM-MIMO frame share the same support, and Eq.(\ref{formula-supp2}) means the sparsity of these vectors are all $N_u$. These two properties determine that the all transmitted signal vectors in the same IM-MIMO frame are jointly sparse and each transmitted signal matrix of an IM-MIMO frame $ \mathbf{X}_{i} = \left[\mathbf{x}_{1,i}, \mathbf{x}_{2,i}, \cdots, \mathbf{x}_{T,i} \right] \in \mathbb{C}^{N_t \times T} $ has the structural sparsity.\par

Assuming that the period to send an IM-MIMO frame is less than the coherence time of the channel, the channel impulse response does not change for transmitting all signal vectors in an IM-MIMO frame. Therefore, we have
\begin{equation}
	\label{formula-channel}
	\mathbf{H}_{1,i}=\mathbf{H}_{2,i}=\cdots=\mathbf{H}_{T,i}=\mathbf{H}_{i},
\end{equation}
where $ \mathbf{H}_i \in \mathbb{C}^{N_r \times N_t } $ is the channel impulse response during the transmission of the $i$-th IM-MIMO frame. Based on Eq.(\ref{formula-mapping}) and Eq.(\ref{formula-channel}), Eq.(\ref{formula-transmission}) can be rewritten as
\begin{equation}
	\label{formula-cs}
	\mathbf{Y}_i = \mathbf{H}_i\mathbf{X}_i + \mathbf{N}_i,
\end{equation}
where $ \mathbf{Y}_i = \left[ \mathbf{y}_{1,i}, \mathbf{y}_{2,i}, \cdots, \mathbf{y}_{T,i} \right] \in \mathbb{C}^{N_r \times T}$ represents the $i$-th received IM-MIMO frame and $ \mathbf{N}_i = \left[ \mathbf{n}_{1,i}, \mathbf{n}_{2,i}, \cdots, \mathbf{n}_{T,i} \right] \in \mathbb{C}^{N_r\times T} $ represents the AWGN for the $i$-th IM-MIMO-OFDM frame. Since the transmitted signal vectors in an IM-MIMO frame are jointly sparse and $ \mathbf{X}_i $ has the property of structural sparsity, the transmission process expressed as Eq.(\ref{formula-cs}) can be considered as a compress sensing process. Accordingly, the detection process at the receiver to recover the transmitted signal can be formulated as a joint sparse reconstruction model (JSRM). In JSRM, $ \mathbf{X}_i $ is the transmitted signal matrix composed of the jointly sparse signal vectors that needs to be recovered, $ \mathbf{H}_i $ is the sensing matrix and $ \mathbf{Y}_i $ is the received signal matrix that contains the measurement vectors. Based on the structure sparsity of $ \mathbf{X}_i $, and each signal vector in $ \mathbf{X}_i $ shares the same sensing matrix (i.e., $ \mathbf{H}_i $), $ \mathbf{X}_i $ can be jointly recovered through $ \mathbf{Y}_i $, which can be formulated as a optimization problem with the consideration of AWGN.
\begin{subequations}\label{formula-JSRM}
	\begin{align}
	\hat{\mathbf{X}}_i &= \mathop{\arg\min}_{\mathbf{X}_i} \ {\parallel supp_{\textcolor{blue}{2}}(\mathbf{X}_i) \parallel}_1 \tag{\ref{formula-JSRM}}\\
	s.t.\  &Eq.(\ref{formula-supp1}), Eq.(\ref{formula-supp2})\\
	&\frac{1}{T}\sum_{j=1}^{T}{\|\mathbf{y}_{j,i}-\mathbf{H}_i{\color{blue}\mathbf{x}}_{j,i}\|}_2^2 < \epsilon
	\end{align}
\end{subequations}
where $supp_{\textcolor{blue}{2}}(\mathbf{X}_i) = supp_{\textcolor{blue}{1}}(\mathbf{x}_{1,i}) \cup supp_{\textcolor{blue}{1}}(\mathbf{x}_{2,i}) \cup \cdots \cup supp_{\textcolor{blue}{1}}(\mathbf{x}_{T,i})$ and  $ \epsilon $ is a predetermined noise level of the system.

There exist many methods to realize the sparse reconstruction, such as greedy based methods like orthogonal matching pursuit (OMP) \cite{CS-OMP} and simultaneous orthogonal matching pursuit (SOMP) \cite{CS-SOMP}, relaxed mixed norm minimization methods and Bayesian methods. However, these methods need the prior knowledge of sensing matrix, which is the channel matrix (i.e., $ \mathbf{H}_i $) in our system model. Considering the fact that perfect channel estimation leads to greater overhead and may be unrealistic in actual wireless communication systems, we introduce DL to realize the signal detection with the consideration of imperfect channel estimation, which will be detailed in the next section.

\section{Deep Learning Based Detection Framework}\label{IMRecoNet}
To be more practical, the impact brought by imperfect channel estimation and the correlation between antennas should be considered at the receiver. Aiming at solving the joint sparse reconstruction problem formulated in Section \ref{problem-formulation} with the consideration of imperfect channel estimation and the correlation between antennas, a DL base detection framework, called IMRecoNet, is designed in this section. Besides, for adapting to the signal represented by complex values in wireless communication systems, complex-valued operations are introduced to the design of IMRecoNet.

\subsection{The architecture of IMRecoNet}\label{IMRecoNet-Arch}
\begin{figure*}[htbp]
	\centering
	\includegraphics[width=0.35\linewidth]{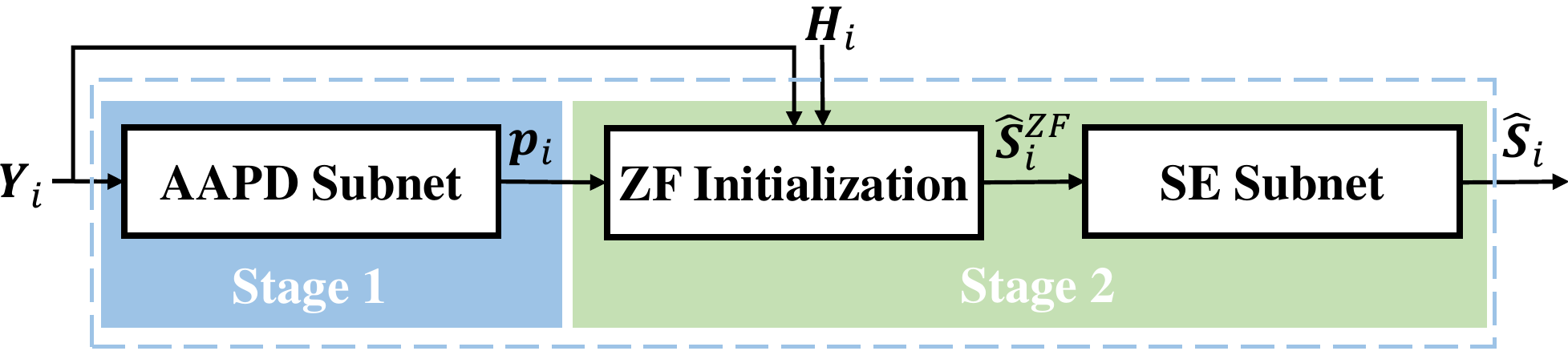}
	\captionsetup{name={Fig.},labelsep=period,singlelinecheck=off,font={small}}
	\caption{The architecture of proposed IMRecoNet.}
	\label{fig:IMRecoNet}
\end{figure*}

According to the system model in Section \ref{system-model}, IM-MIMO systems transmit information through TAC $J_i$ and actually transmitted signal vector $\mathbf{s}_{j,i}$. Therefore, we can decompose the joint sparse reconstruction problem formulated in Section \ref{problem-formulation} into two steps. The first step is to estimate the position of non-zero elements (i.e., the TAC $J_i$), and the second step is to estimate the specific value of the non-zero elements (i.e., the actually transmitted signal matrix $\mathbf{S}_i = \left[\mathbf{s}_{1,i},\mathbf{s}_{2,i},\cdots,\mathbf{s}_{T,i}\right] \in \mathbb{C}^{N_u \times T}$). Based on $J_i$ and $\mathbf{S}_i$, the transmitted signal matrix $ \mathbf{X}_i $ for the $i$-th IM-MIMO frame can be reconstructed easily. We design a DL based signal detection framework, called IMRecoNet, to realize the two steps mentioned before, whose architecture is shown in Fig.\ref{fig:IMRecoNet}. IMRecoNet consists of two CNNs and a traditional zero-forcing (ZF) detector. At the first stage, the AAP detection (AAPD) subnet is utilized to detect the activation status of each TA. At the second stage, ZF detector combined with signal enhancement (SE) subnet realize to estimate the actually transmitted signal based on the activation information of TAs predicted by AAPD subnet. It is worth noting that both AAPD subnet and SE subnet are running in an end-to-end manner, which means they do not need CSI at the training and detection period.

\begin{figure*}[htbp]
	\centering	
\subfigure[AAPD]{	
	\includegraphics[width=0.35\linewidth]{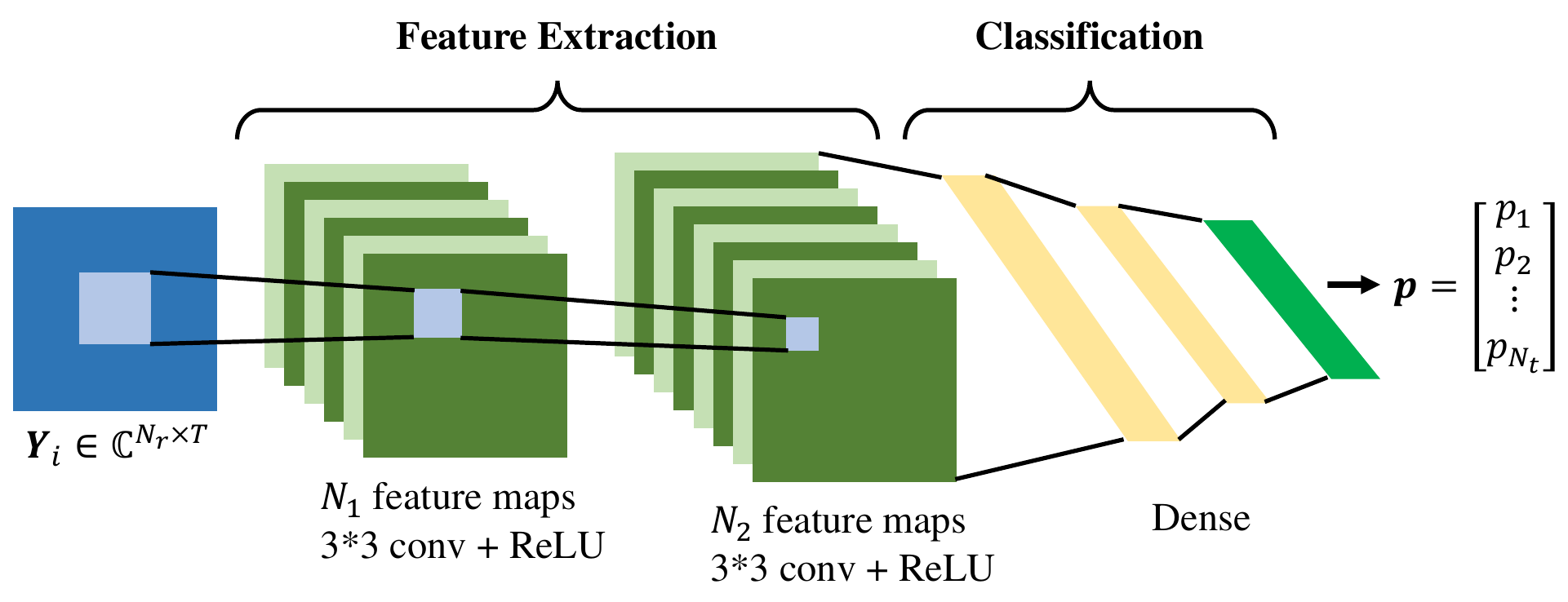}
    \label{fig:AAPD}}
\subfigure[SE]{	
	\includegraphics[width=0.35\linewidth]{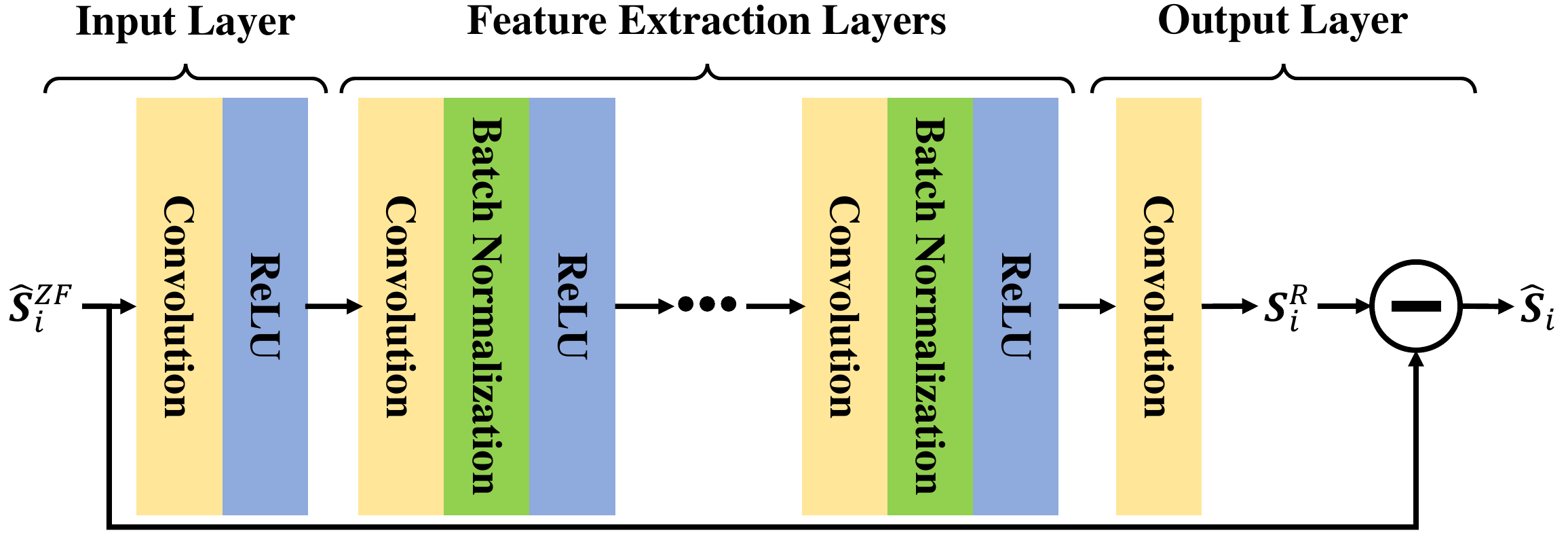}
    \label{fig:SE}}
\captionsetup{name={Fig.},labelsep=period,singlelinecheck=off,font={small}}
\caption{The architecture of AAPD and SE subnet.}
\end{figure*}

In fact, the signals transmitted by different TAs pass through different wireless channels to reach the receiver. Therefore, the ``fingerprint" information of the wireless channel is hidden in the received signal, and it also contains the activation state information of each TA. Based on this feature hidden in the received signal, the data-driven AAPD subnet can mine this internal feature well and predict the activation probability of all TAs at once. The architecture of AAPD subnet is shown in Fig.\ref{fig:AAPD}.

As shown in Fig.\ref{fig:AAPD}, the input of AAPD subnet is the received signal matrix for an IM-MIMO frame and then two convolutional layers are used for feature extraction. In order to speed up the convergence of the network, two convolutional layers use ReLU as the activation function. The prediction of the activation state of each TA is actually a two-class problem (i.e., active or inactive). Therefore, three full-connection layers are used to map features extracted by the convolution layers, and Sigmoid function is adopted as the activation function of the output layer to give the activation probability of each TA. Since AAPD subnet solves a classification problem, we use binary cross-entropy function to train it, which is expressed as
\begin{equation}
 \begin{aligned}
	\label{form:aapd_loss}
	L_{AAPD}(\boldsymbol{\theta}_1)&=\frac{1}{N_t}\sum_{m=1}^{N_t}\left[\mathbf{g}_i(m)\log(\mathbf{p}_i(m))\right]\\
	&+\sum_{m=1}^{N_t}\left[(1-\mathbf{g}_i(m))\log(1-\mathbf{p}_i(m))\right],
\end{aligned}
\end{equation}
where $\boldsymbol{\theta}_1$ denotes the parameters of AAPD subnet, $\mathbf{g}_i$ is the AAP for the $i$-th IM-MIMO frame defined in Eq.(\ref{form:aap}), and $ \mathbf{p}_i = \left[ p_{1,i}, p_{2,i}, \cdots, p_{N_t,i} \right] $ represents the activation probability vector of TAs for the $i$-th IM-MIMO frame. The entire workflow of AAPD subnet can be expressed as
\begin{align}
&\mathbf{p}_i = f_{AAPD}(\mathbf{Y}_i;\boldsymbol{\theta}_1),\label{formula-aap3} \\
&\hat{J}_i = \left\{ n\ having \ the\ top\  N_u\ largest\ \mathbf{p}_i(n) \right\},\label{formula-aap4}
\end{align}
where $ f_{AAPD}(\cdot) $ represents the function of AAPD subnet, $ \hat{J}_i $ is the estimated TAC for the $i$-th IM-MIMO frame and $ n \in \left\{ 1, 2, \cdots, N_t \right\} $ is the index of TAs. 

Since the inactive TAs actually do not make any contribution to the received signal, when $J_i$ is obtained through the AAPD subnet, Eq.(\ref{formula-cs}) can be rewritten as
\begin{equation}
	\label{form:small-cs}
	\mathbf{Y}_i = \mathbf{H}_i(\hat{J}_i)\mathbf{S}_i+\mathbf{N}_i,
\end{equation}
where $\mathbf{H}_i(\hat{J}_i) \in \mathbb{C}^{N_r \times N_u}$ consists of the columns of $\mathbf{H}_i$ according to the elements of $J_i$, and $\mathbf{S}_i = \left[\mathbf{s}_{1,i},\mathbf{s}_{2,i},\cdots,\mathbf{s}_{T,i}\right] \in \mathbb{C}^{N_u \times T}$ is the actually transmitted signal matrix. For Eq.(\ref{form:small-cs}), we can obtain the initial estimation through ZF detector, which can be expressed as
\begin{equation}
\label{formula-ls}
\mathbf{\hat{S}}_i^{ZF} = \mathbf{W}_i^{ZF}\mathbf{Y}_i={\left[ {\mathbf{H}_i(\hat{\mathbf{J}}_i)}^H{\mathbf{H}_i(\hat{\mathbf{J}}_i)} \right]}^{-1}{\mathbf{H}_i(\hat{\mathbf{J}}_i)}^H{\mathbf{Y}_i},
\end{equation}
where $\mathbf{W}_i^{ZF}={\left[ {\mathbf{H}_i(\hat{\mathbf{J}}_i)}^H{\mathbf{H}_i(\hat{\mathbf{J}}_i)} \right]}^{-1}{\mathbf{H}_i(\hat{\mathbf{J}}_i)}^H $ is the interference cancellation matrix. Considering the  noise enhancement introduced by ZF detector and channel estimation error, we further introduce SE subnet in the second stage to enhance $\mathbf{\hat{S}}_i^{ZF}$ estimated by ZF detector through Eq.(\ref{formula-ls}), whose architecture is shown in Fig.\ref{fig:SE}.


As shown in Fig.\ref{fig:SE}, SE subnet use residual learning \cite{he2016deep} to enhance $\mathbf{\hat{S}}_i^{ZF}$. SE subnet is composed of input layer, feature extraction layer and output layer. The input of the SE subnet is the rough estimation result of the ZF detector, and the output is the final detection result after enhancement, which is essentially a fitting problem. Therefore, the SE subnet uses the mean square error (MSE) as its loss function:
\begin{equation}
	\label{form:mse}
	L_{SE}(\boldsymbol{\theta}_2)=\frac{1}{N}\sum_{i=1}^{N}\|f_{SE}(\hat{S}_i^{ZF};\boldsymbol{\theta}_2)-\mathbf{S}_i\|_2^2,
\end{equation}
where $\boldsymbol{\theta}_2$ denotes the parameter set of SE subnet, and $f_{SE}(\cdot)$ represents the function of SE subnet.

\subsection{Complex-valued IMRecoNet}\label{CVCNN}
The mainstream DL libraries (such as Tensorflow \cite{tensorflow} and PyTorch \cite{pytorch}) are designed for processing real-valued data, which is inconsistent with the processing of complex-valued signals in wireless communication systems. In order to improve the ability of IMRecoNet to process complex-valued signals, we adopt complex-valued operations into AAPD subnet and SE subnet.

\emph{\textbf{1) Complex-Valued \textcolor{blue}{Differentiation}:}} When training neural networks, the backpropagation algorithm needs to calculate gradient to update weight parameters. However, the introduction of complex values limit the gradient calculation of the activation function and loss function during backpropagation. Wirtinger calculus \cite{kreutz2009complex} provides an alternative method for calculating the \textcolor{blue}{differentiation} of a complex function, which is defined as
\begin{equation}
	\begin{small}
	\label{form:complex_gradient1}
	\frac{\partial f}{\partial z}  \triangleq \frac{1}{2}(\frac{\partial f}{\partial x}-j\frac{\partial f}{\partial y}), 
    \end{small}
\end{equation}

\begin{equation}
	\label{form:complex_gradient2}
	\frac{\partial f}{\partial \bar{z}}  \triangleq \frac{1}{2}(\frac{\partial f}{\partial x}+j\frac{\partial f}{\partial y}).
\end{equation}
Eq.(\ref{form:complex_gradient1}) and Eq.(\ref{form:complex_gradient2}) are called as $\mathbb{R}$-derivative and conjugate $\mathbb{R}$-derivative, respectively. Based on Eq.(\ref{form:complex_gradient1}) and Eq.(\ref{form:complex_gradient2}), researchers in \cite{amin2011wirtinger} and \cite{hirose6515223} define the complex gradient as
\begin{equation}
	\label{form:complex_gradient_definition}
	\nabla_zf=2\frac{\partial f}{\partial \bar{z}}.
\end{equation}
Although Tensorflow does not directly support the implementation of complex values, its automatic differentiation algorithm supports the use of complex values \cite{hoffmann2016hitchhiker}. For $f:\mathbb{C}\rightarrow\mathbb{C}$, Tensorflow calculates the complex gradient as
\begin{equation}
	\label{form:tensorflow_autodiff}
	\nabla_zf=\overline{\frac{\partial f}{\partial z}+\frac{\partial \bar{f}}{\partial z}}.
\end{equation}
It is straightforward to prove that Eq.(\ref{form:complex_gradient_definition}) and Eq.(\ref{form:tensorflow_autodiff}) are equivalent based on the characteristics in \cite{kreutz2009complex}.

For any real-valued loss function $\mathcal{L}:\mathbb{C}\rightarrow\mathbb{R}$ and any complex-valued function $g:\mathbb{C}\rightarrow\mathbb{C}$, the complex gradient can be obtained through chain rule as

\begin{equation}
	\label{form:chain}
	\frac{\partial \mathcal{L} \circ g}{\partial \bar{z}} = \frac{\partial \mathcal{L}}{\partial g}\frac{\partial g}{\partial \bar{z}} +\frac{\partial \mathcal{L}}{\partial \bar{g}}\frac{\bar{g}}{\partial \bar{z}}.
\end{equation}

Eq.(\ref{form:complex_gradient_definition}) and Eq.(\ref{form:chain}) make it possible to train complex-valued neural networks.

\emph{\textbf{2) Complex-Valued Convolution Layer:}} Assuming that the $l$-th complex-valued convolution layer outputs $ n $ complex feature maps (denoted as $ \mathbf{M}^l_{1\sim n} $), we can expand the $ n $ complex feature maps to $ 2n $ real feature maps as: $\mathbf{\widetilde{M}}^l_{1 \sim n} = \Re\left\{\mathbf{M}^l_{1\sim n}\right\}$ and $\mathbf{\widetilde{M}}^l_{n+1 \sim 2n} = \Im\left\{\mathbf{M}^l_{1\sim n}\right\}$. Expanding the $m$ complex convolution kernels $\mathbf{W}^{l+1}_{1 \sim m}$ of the $(l+1)$-th convolution layer into $2m$ real convolution layers (i.e., ${\mathbf{\widetilde{W}}}^{l}_{1 \sim m}$ and ${\mathbf{\widetilde{W}}}^{l}_{m+1 \sim 2m}$ ) in the same way, the complex valued convolution layer can be defined as 

\begin{equation}
\begin{aligned}
	\label{formula-4.3}
	\mathbf{\widetilde{M}}^{l+1}_{1 \sim m} &= \Re\left\{\mathbf{W}^{l+1}_{1 \sim m}\circledast\mathbf{M}^l_{1\sim n}\right\} \\
	 &=\mathbf{\widetilde{W}}^{l+1}_{1 \sim m}\mathbf{\widetilde{M}}^{l}_{1 \sim n} - \mathbf{\widetilde{W}}^{l+1}_{m+1 \sim 2m}\mathbf{\widetilde{M}}^{l}_{n+1 \sim 2n},	
\end{aligned}
\end{equation}

\begin{equation}
\begin{aligned}
	\label{formula-4.4}
	\mathbf{\widetilde{M}}^{l+1}_{m+1 \sim 2m} &= \Im\left\{\mathbf{W}^{l+1}_{1 \sim m}\circledast\mathbf{M}^l_{1\sim n}\right\} \\
	&=\mathbf{\widetilde{W}}^{l+1}_{m+1 \sim 2m}\mathbf{\widetilde{M}}^{l}_{1 \sim n} + \mathbf{\widetilde{W}}^{l+1}_{1 \sim m}\mathbf{\widetilde{M}}^{l}_{n+1 \sim 2n},
\end{aligned}
\end{equation}
 where $ \circledast $ represents convolution operation. Combining $\mathbf{\widetilde{M}}^{l+1}_{1 \sim m}$ and $\mathbf{\widetilde{M}}^{l+1}_{m+1 \sim 2m}$ can get the final output of the $(l+1)$-th complex convolution layer (i.e., $\mathbf{{M}}^{l+1}_{1 \sim m}$).

\emph{\textbf{3) Complex-Valued Activation Function:}} For the complex activation function, we extend ReLU and Sigmoid function through following expression to make them support complex values.
\begin{align}
	f_1(z)&=Sigmoid(\Re\left\{z\right\})+jSigmoid(\Im\left\{z\right\})\label{form:complex_sigmod},\\
	f_2(z)&=ReLU(\Re\left\{z\right\})+jReLU(\Im\left\{z\right\})\label{form:complex_relu}.
\end{align}
\textcolor{blue}{\subsection{Batch Normalization of Complex Value}
In order to avoid the potential over-fitting hazards, we use the batch normalization for the complex number to accelerate the convergence speed of the neural network, and prevent the gradient disappearance as well as control explosion. Different from the real number, complex number has real and imaginary parts. If the complex numbers are only scaled and shifted to make the mean to $0$ and the variance to $1$ as real numbers, there is no guarantee that real part and imaginary part have the same variance, and the distribution of the normalized value may be offset. For normalized complex domain, it can be regarded as a two-dimensional matrix whiten problem, by using two main components of the square root of the variance of each component to scale the value. This can be done by a covariance matrix, calculated as follows:
	\begin{align}
		\widetilde{\mathbf{x}}=(\mathbf{V})^{-\frac{1}{2}}(\mathbf{x}-\mathbb{E}\{\mathbf{x}\})
	\end{align}
where $\mathbf{V}$ is the covariance matrix with the definition 
	\begin{small}
\begin{align}
	\mathbf{V} &=
	\left(
	\begin{array}{cc}   
		V_{rr} & V_{ri} \\  
		V_{ir} & V_{ii} \\  
	\end{array}
	\right) 
	  =
	\left(
	\begin{array}{cc}   
		Cov(\Re(\mathbf{x}),\Re(\mathbf{x})) & Cov(\Re(\mathbf{x}),\Im(\mathbf{x})) \nonumber \\  
		Cov(\Im(\mathbf{x}),\Re(\mathbf{x})) & Cov(\Im(\mathbf{x}),\Im(\mathbf{x})) \\  
	\end{array}
	\right)
\end{align}
	\end{small}
The above normalization procedure eliminates the relevance of the imaginary part and real part in complex number and avoids the real part and imaginary part in adjusting coupling, which reduces over-fitting risk in the complex neural network\cite{overfitting}. 
Similar to the batch normalization of real number, the complex batch normalization also uses two parameters  $\gamma$ and $\beta$ to scale and translate the normalized complex number, which is shown in the following type:
\begin{align}
	\mathbb{C}BN=\mathbf{\gamma}\mathbf{\widetilde{x}}+\mathbb\beta
\end{align}
$\gamma$ is a matrix with four parameters, which is given as 
$$      
\gamma=
\left(                 
\begin{array}{cc}   
	\gamma_{rr} & \gamma_{ri} \\  
	\gamma_{ir} & \gamma_{ii} \\  
\end{array}
\right)                 
$$
  $\beta$ is a complex vector with the same dimension as $\mathbf{\widetilde{x}}$. 
   In order to make the output of complex batch normalization obey the standard Gaussian distribution, both $\gamma_{rr}$ and $\gamma_{ir}$  need to be initialized to $\frac{1}{\sqrt{2}}$. Besides, the real part and the imaginary part of the complex offset $\beta$  need to be initialized to 0.}
\subsection{Training and Deployment IMRecoNet}\label{CVCNN-AAPD}
Since IMRecoNet contains two independent CNNs, we use a step-by-step training method to train IMRecoNet. Based on the generated training dataset, we first train the AAPD subnet. Based on the trained AAPD subnet, we generate the rough estimation result $\mathbf{S}^{ZF}_i$ of the transmitted signal matrix $\mathbf{S}_i$ according to Eq.(\ref{formula-ls}). Finally, we use the rough estimation result of the transmitted signal matrix $\mathbf{S}^{ZF}_i$ and the actual transmitted signal matrix $\mathbf{S}_i$ to train the SE subnet. The entire training process is shown in Algorithm \ref{algo:IMRecoNet_train}.

\begin{algorithm}[htbp]
	\small
	\SetAlgoLined
	\textbf{Input:} $\left\{\mathbf{g}_i\right\}_{i=1}^{N}$: AAP set; $\left\{\mathbf{S}_i\right\}_{i=1}^{N}$: actually transmitted signal matrix; $\left\{\mathbf{Y}_i\right\}_{i=1}^{N}$: received signal matrix; $\left\{\mathbf{H}_i\right\}_{i=1}^{N}$: channel matrix; $\gamma_1, \gamma_2$: error threshold \\
	\textbf{Output:} $\boldsymbol{\Theta}=\left\{\boldsymbol{\theta}_1,\boldsymbol{\theta}_2\right\}$: parameter set of IMRecoNet\;
	
	initialize parameters $\boldsymbol{\theta}_1$\;
	\While{$error \geqslant \gamma_1$}{
		use dataset $\left\{\mathbf{Y}_i,\mathbf{g}_i\right\}_{i=1}^{N}$ to train AAPD subnet\;
		minimize Eq.(\ref{form:aapd_loss}), update paremeters $\boldsymbol{\theta}_1$\;
	}
	get $\mathbf{p}_i$ through Eq.(\ref{formula-aap3})\;
	get $\hat{J}_i$ through Eq.(\ref{formula-aap4})\;
	get $\mathbf{S}^{ZF}_i$ through Eq.(\ref{formula-ls})\;
	initialize paremeters $\boldsymbol{\theta}_2$\;
	\While{$error \geqslant \gamma_2$}{
		use dataset $\left\{\mathbf{S}^{ZF}_i, \mathbf{S}_i\right\}_{i=1}^{i=N}$ to train SE subnet\;
		minimize Eq.(\ref{form:mse}), update parameters $\theta_2$\;
	}
	\textbf{Return:} {$\boldsymbol{\Theta}=\left\{\boldsymbol{\theta}_1,\boldsymbol{\theta}_2\right\}$}\;
	\caption{Training IMRecoNet}
	\label{algo:IMRecoNet_train}
\end{algorithm}

When the optimal parameters set $\boldsymbol{\Theta}=\left\{\boldsymbol{\theta}_1,\boldsymbol{\theta}_2\right\}$ is obtained, the signal detection process of IMRecoNet can be expressed as
\begin{equation}
\begin{aligned}
	\label{form:IMNet}
	<\hat{\mathbf{J}}_i, \hat{\mathbf{S}}_i>&=f_{IMRecoNet}(\mathbf{Y}_i; \boldsymbol{\Theta})\\
	&=f_{SE}(f_{ZF}(f_{AAPD}(\mathbf{Y}_i;\boldsymbol{\theta}_1),\mathbf{H}_i,\mathbf{Y}_i);\boldsymbol{\theta}_2)
\end{aligned}
\end{equation}
where $f_{IMRecoNet}(\cdot)$ and $f_{SE}(\cdot)$ are the function represented by the IMRecoNet and ZF detector, respectively. First, the AAPD subnet predicts the activation probability of each TA based on the received signal matrix $\mathbf{Y}_i$. The antenna activation vector $\mathbf{p}_i$ predicted by the AAPD subnet will be mapped to the TAC $\hat{J}_i$. After that, according to Eq.(\ref{formula-ls}), using the the received signal matrix $\mathbf{Y}_i$, the channel matrix $\hat{\mathbf{H}}_i$ obtained by the receiving end and the TAC $\hat{J}_i$ predicted in the previous step, the rough estimation result of the ZF detector $\mathbf{S}^{ZF}_i$ can be calculated. Finally, the SE subnet is used to enhance the rough estimation result of the ZF detector to obtain the final signal detection result $\hat{\mathbf{S}}_i$. The signal detection process of IMRecoNet is shown in Algorithm \ref{algo:IMNet_Det}.

\begin{algorithm}[htbp]
	\small
	\SetAlgoLined
	\textbf{Input:} $\mathbf{Y}_i$: the received signal matrix; $\mathbf{H}_i$: channel matrix\;
	\textbf{Output:} $\hat{J}_i$: the predicted TAC; $\hat{\mathbf{S}}_i$: the actually transmitted signal matrix\;
	
	load IMRecoNet according to $\boldsymbol{\Theta}=\left\{\boldsymbol{\theta}_1,\boldsymbol{\theta}_2\right\}$\;
	\While{receive $\mathbf{Y}_i$}{
		predict activation probability of each TA: $\mathbf{p}_i=f_{AAPD}(\mathbf{Y}_i;\boldsymbol{\theta}_1)$\;
		map $\mathbf{p}_i$ to $\hat{J}_i$ $\mathbf{p}_i\rightarrow\hat{J}_i$\;
		calculate $\hat{\mathbf{S}}_i^{ZF}$ according to Eq.(\ref{formula-ls})\;
		enhance $\hat{\mathbf{S}}_i^{ZF}$: $\hat{\mathbf{S}}_i=f_{SE}(\hat{\mathbf{S}}_i^{ZF};\boldsymbol{\theta}_2)$\;
	}
	\textbf{Return:} $\hat{J}_i,\hat{\mathbf{S}}_i$ \;
	\caption{Signal Detection Process of IMRecoNet}
	\label{algo:IMNet_Det}
\end{algorithm}

\subsection{Complexity Analysis}
The number of parameters of a deep learning model is one of the important indicators to measure the complexity of the model, and it also affects the convergence speed of the model. For IMRecoNet, whether using real-valued or complex-valued operations, the main modules are convolution layer, full-connection layer and batch normalization layer. Among them, the batch normalization layer has fewer parameters and accounts for a relatively small proportion of the weight parameters of the entire IMRecoNet. Therefore, the number of weight parameters of IMRecoNet mainly comes from the convolution layer and the full-connection layer. The theoretical calculation results of the number of weight parameters of the convolution layer and the full-connection layer are shown in Table \ref{tab:IMRecoNet_ConvParameters} and Table \ref{tab:IMRecoNet_FullConnParameters}, respectively.

For a real-valued convolution layer with $N_{in}$ feature maps and $N_{out}$ feature maps, the kernel size is $k \times k \times N_{in}$, therefore, one convolution layer contains $k \times k \times N_{in} \times N_{out}$ parameters. When using complex-valued convolution, every two real-valued feature maps correspond to a complex-valued feature map, i.e., $\frac{N_{in}}{2}$ ($\frac{N_{out}}{2}$) complex-valued feature maps consist of $\frac{N_{in}}{2}$ ($\frac{N_{out}}{2}$) real parts and $\frac{N_{in}}{2}$ ($\frac{N_{out}}{2}$) imaginary parts. Complex-valued convolution kernels also have real parts and imaginary parts. According to the rules of complex-valued convolution operation expressed in Eq.(\ref{formula-4.3}) and Eq.(\ref{formula-4.4}), it can be known that the dimensions of the real and imaginary convolution kernels are both $k \times k \times \frac{N_{in}}{2}$. Since the real part feature maps and the imaginary part feature maps share the same real part convolution kernels and the imaginary part convolution kernels, according to Eq.(\ref{formula-4.3}) and Eq.(\ref{formula-4.4}), there exist $k \times k \times \frac{N_{in}}{2} \times \frac{N_{out}}{2}$ real part parameters and $k \times k \times \frac{N_{in}}{2} \times \frac{N_{out}}{2}$ imaginary part parameters, i.e., a total of $\frac{k \times k \times N_{in} \times N_{out}}{2}$ parameters, for connecting $\frac{N_{in}}{2}$ input complex-valued feature maps to $\frac{N_{out}}{2}$ output complex complex-valued feature maps. Compared with real-valued convolution, complex-valued convolution cut the number of parameters by half, greatly reduces the total number of IMRecoNet's parameters.

\begin{table*}[htbp]
	\centering\small
	\caption{The number of parameters of real-valued and complex-valued convolution layer.}
	\label{tab:IMRecoNet_ConvParameters}
	\begin{tabular}{c|c|c|c|c}
		\hline\hline
		& Input Channels      & Output Channels      & Kernel Size         & Parameters(\textcolor{blue}{Complexity})               \\
		\hline
		Real-Valued                  & $N_{in}$              & $N_{out}$             & $k \times k$                & $ k \times k \times N_{in} \times N_{out} $     \\
		\hline
		\multirow{4}{*}{Complex-Valued} & \multirow{2}{*}{$ N_{in} / 2 $ real parts} & \multirow{2}{*}{$ N_{out} / 2 $ real parts} & \multirow{4}{*}{$ k \times k $} & \multirow{4}{*}{$\left(k \times k \times N_{in} \times N_{out}\right) \ / \ 2$} \\
		&                    &                    &                    &                    \\
		& \multirow{2}{*}{$ N_{in} / 2  $ imaginary parts} & \multirow{2}{*}{$ N_{out} / 2 $ imaginary parts} &                    &                    \\
		&                    &                    &                    &                    \\
		\hline\hline
	\end{tabular}
\end{table*}

\begin{table*}[htbp]
	\centering\small
	\caption{The number of parameters of real-valued and complex-valued full-connection layer.}
	\label{tab:IMRecoNet_FullConnParameters}
	\begin{tabular}{c|c|c|c}
		\hline\hline
		& Input Dimension     & Output Dimension        & Parameters(\textcolor{blue}{Complexity})             \\ \hline
		Real-Valued                  & $N_{in}$                  & $N_{out}$                  & $N_{in} \times N_{out}$                  \\ \hline
		\multirow{4}{*}{Complex-Valued} & \multirow{2}{*}{$N_{in} / 2$ real parts} & \multirow{2}{*}{$N_{out} / 2$ real parts} & \multirow{4}{*}{$\left(N_{in} \times N_{out}\right) \ / \ 2$} \\
		&                    &                    &                    \\
		& \multirow{2}{*}{$N_{in} / 2$ imaginary parts} & \multirow{2}{*}{$N_{out} / 2$ imaginary parts} &                    \\
		&                    &                    &                    \\
		\hline\hline
	\end{tabular}
\end{table*}

For the real-valued full-connection layer with input dimension $N_{in}$ and output dimension $N_{out}$, the number of parameters is $N_{in} \times N_{out}$. In the complex-valued full-connection layer, every two real numbers correspond to a complex number, so the input and output dimensions will be halved, that is, the input dimension becomes $\frac{N_in}{2}$, and the output dimension becomes $\frac{N_{out}}{2}$. The operation process of the complex-valued full-connected layer is similar to that of the complex-valued convolutional layer. The number of real and imaginary parameters required to connect the $N_{in}$-dimensional input complex-valued vector to the $N_{out}$-dimensional output complex-valued vector is both $\frac{N_{in}}{2} \times \frac{N_{out}}{2}$, the total number of parameters is $\frac{N_{in} \times N_{out}}{2}$, which is only half of the number of parameters of real-valued full-connection layer.

From the above theoretical analysis, it can be seen that whether it is a convolutional layer or a full-connection layer, the number of parameters based on the complex-valued operation is only half of that based on the real-valued operation. IMRecoNet mainly consists of convolutional layers and full-connection layers. Therefore, the number of parameters of complex-valued IMRecoNet is only about half that of the real-valued IMRecoNet. The smaller number of parameters, on the one hand, ensures the convergence speed of the complex-valued IMRecoNet during training, and on the other hand, it ensures that it requires less computing resources and computing resources at runtime. 
A detailed comparison will be given in the subsequent experimental analysis part about the specific number of weight parameters of the complex-valued IMRecoNet and the real-valued IMRecoNet. \textcolor{blue}{Besides, the performance of the number of floating point operations (FLOPs) in the signal detection of IMRecoNet, which is widely used in  deep learning process, will be provided in the simulation results to help illustrate the complexity of proposed method.}

\section{Performance Evaluation}\label{Performance-Evaluation}

In this section, we evaluate and discuss the performance of proposed IMRecoNet. Section \ref{Implementation} gives the details of the implementation of the proposed IMRecoNet and several configurations of simulation. Section \ref{Results} demonstrates the simulation results under various scenarios to show the adaptability and robustness of the proposed IMRecoNet.

\subsection{Implementation}\label{Implementation}

Both two subnets (i.e., AAPD subnet and SE subnet) in IMRecoNet are developed through Keras \cite{kerasWeb} with Tensorflow as backend. The two subnets are trained using gradient based Adam method on the Nvidia GTX 1080ti GPU environment. The learning rate and batch size are set to 0.001 and 100, respectively. When we generate training data and test data, the SNR is defined as
\begin{equation}
	SNR(dB)=10\log\frac{\mathbb{E}\left\{\|\mathbf{H}_i\mathbf{x}_{j,i}\|_2^2\right\}}{\mathbb{E}\left\{\|\mathbf{n}_{j,i}\|_2^2\right\}},
\end{equation} 
to measure the noise level. In order to fully verify the performance of IMRecoNet, IMRecoNet is trained and tested at each SNR of {5dB, 10dB, 15dB, 20dB and 25dB}.
\textcolor{blue}{We generate 100,000 sets of data through MATLAB R2020b for large number of antennas scenario, of which  training, validation and testing  datasets accounted for $60\%$,$20\%$ and $20\%$, respectively.}
To verify the adaptability of IMRecoNet, both Rayleigh fading MIMO channel and correlated MIMO channel are considered in our simulations:

\emph{1) Rayleigh fading MIMO channel:} The elements of channle matrix $ \bm{H} $ obeys complex Gaussian distribution, i.e., $ h_{i,j} \sim \mathcal{CN}(0,\frac{1}{N_r}) $.

\emph{2) Correlated MIMO channel:} Based on the Rayleigh fading MIMO channel, we can generate the correlated MIMO channel according to Kronecker model, which can be expressed as
\begin{equation}\label{corrMIMOFormulation}
	\mathbf{H}_c = \mathbf{R}^{\frac{1}{2}}_r\mathbf{H}\mathbf{R}^{\frac{1}{2}}_t,
\end{equation}
where $ \mathbf{H} $ is the independent identical distributed (i.i.d.) Rayleigh fading MIMO channel, $ \mathbf{R}_r $ and $ \mathbf{R}_t $ are the spatial correlation matrix of receive antennas and transmit antennas, respectively. $ \mathbf{R}_r $ and $ \mathbf{R}_t $ are generated through expontial correlation model \cite{corrH} with the same correlation coefficient $\rho$. Specifically, the elements of $ \mathbf{R}_r $ and $ \mathbf{R}_t $ satisfy the following relationship:
\begin{equation}
	\lambda_{i,j} = 
	\begin{cases}
		\rho^{j-i}, & i \leqslant j \\
		\lambda_{j,i}^*, & i > j
	\end{cases}
\end{equation}

In order to demonstrate the robustness of the IMRecoNet, imperfect channel estimation at the receiver is also considered in our simulations. Considering the ML channel estimation \cite{Imperfect-CSIR}\cite{CE-ML}, the imperfect channel estimation can be expressed as 
\begin{equation}
	\label{formula-5.1}
	\bm{\hat{H}} = \bm{H} + \Delta\bm{H},
\end{equation} 
where $ \bm{\hat{H}} $ and $ \Delta\bm{H} $ denote the estimated channel matrix and estimation error, respectively. $ \Delta\bm{H} $ obeys zero mean i.i.d. complex Gaussian distribution with $ \mathbb{E}\left[ {|\Delta{h}_{m,n}|}^2 \right] = \frac{N_t{\sigma}^2_z}{{N_p}{E_p}} $ \cite{Imperfect-CSIR}, where $ N_p $ and $ E_p$ represent the number and the power of pilot symbols, respectively.\par

We compare IMRecoNet with two traditional algorithms:\par
(1) Maximum Likelihood (ML) \cite{SM-ML}\cite{OFDM-IM-ML}: An exhaustive search algorithm  jointly estimates the TAC and actually transmitted signal vector, which exhibits the best BER performance. \par
(2) Simultaneous Orthogonal Matching Pursuit (SOMP) \cite{tropp2006algorithms} : A classic algorithm in the field of compressed sensing, is a fast convergence algorithm based on greedy method.\par

For further verifying the effectiveness and robustness of the proposed IMRecoNet, we carry out simulations in different scenarios, whose details are shown in Table \ref{table-2}. From the table, we can find that, as the number of TAs (i.e., $N_t$) and active TAs (i.e., $N_u$) increases, legal TAC size shows exponential growth, which means that the difficulty and complexity of signal detection increase exponentially.

\begin{table}[htbp]
	\centering
	\normalsize
	\caption{Configurations of IM in different scenarios.}
	\label{table-2}
	\begin{tabular}{cccc}
		\hline\hline
		& $ N_t $ & $ N_u $ & TAC Table Size \\ \hline
		Scenario 1 & 4  & 1 & 4        \\
		Scenario 2 & 16 & 4 & \textcolor{blue}{1024}      \\ \hline\hline
	\end{tabular}
\end{table}

\begin{figure*}
	\centering
	\subfigure[4QAM]{
		\includegraphics[width=0.4\textwidth]{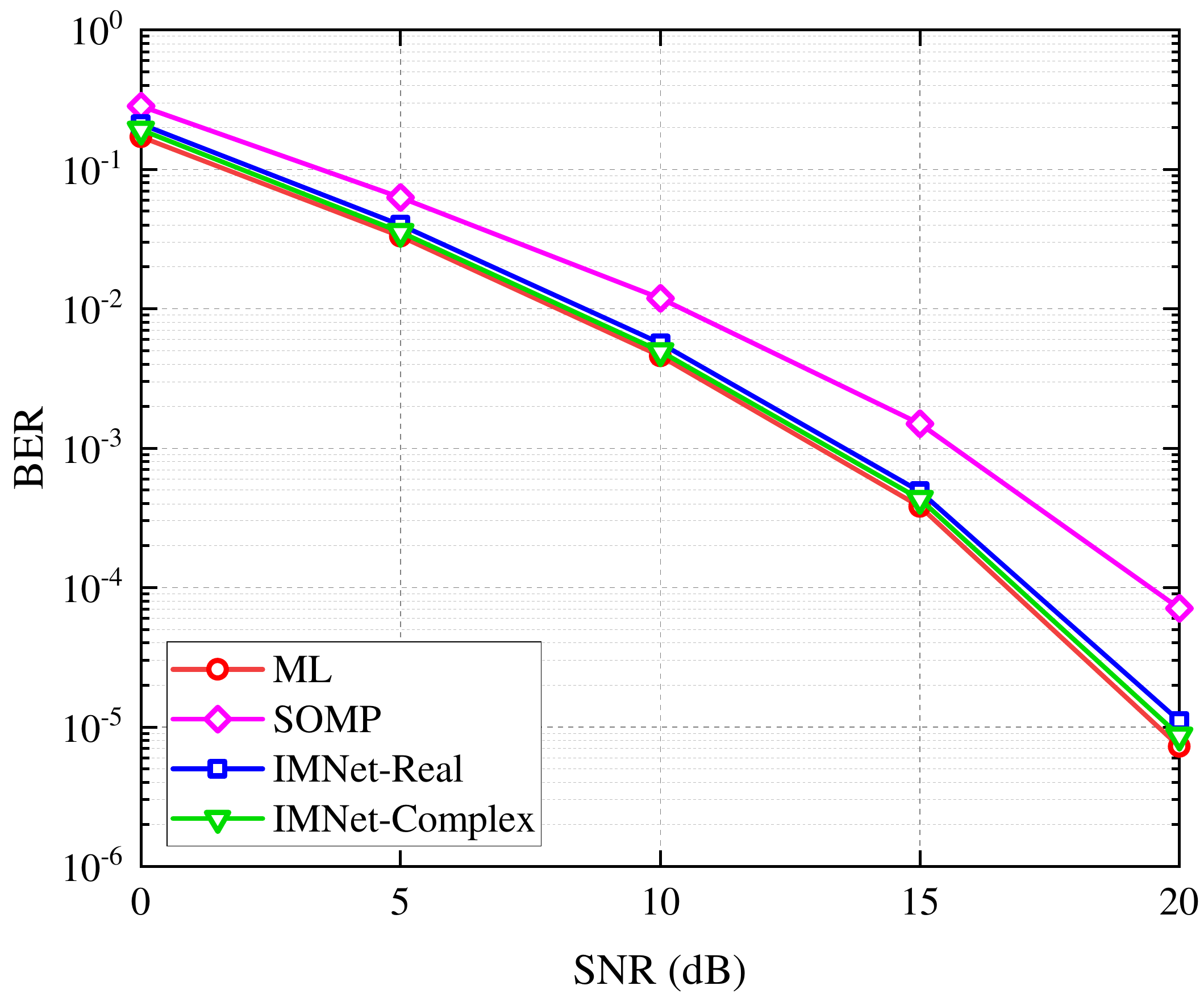}
	}
	\subfigure[16QAM]{
		\includegraphics[width=0.4\textwidth]{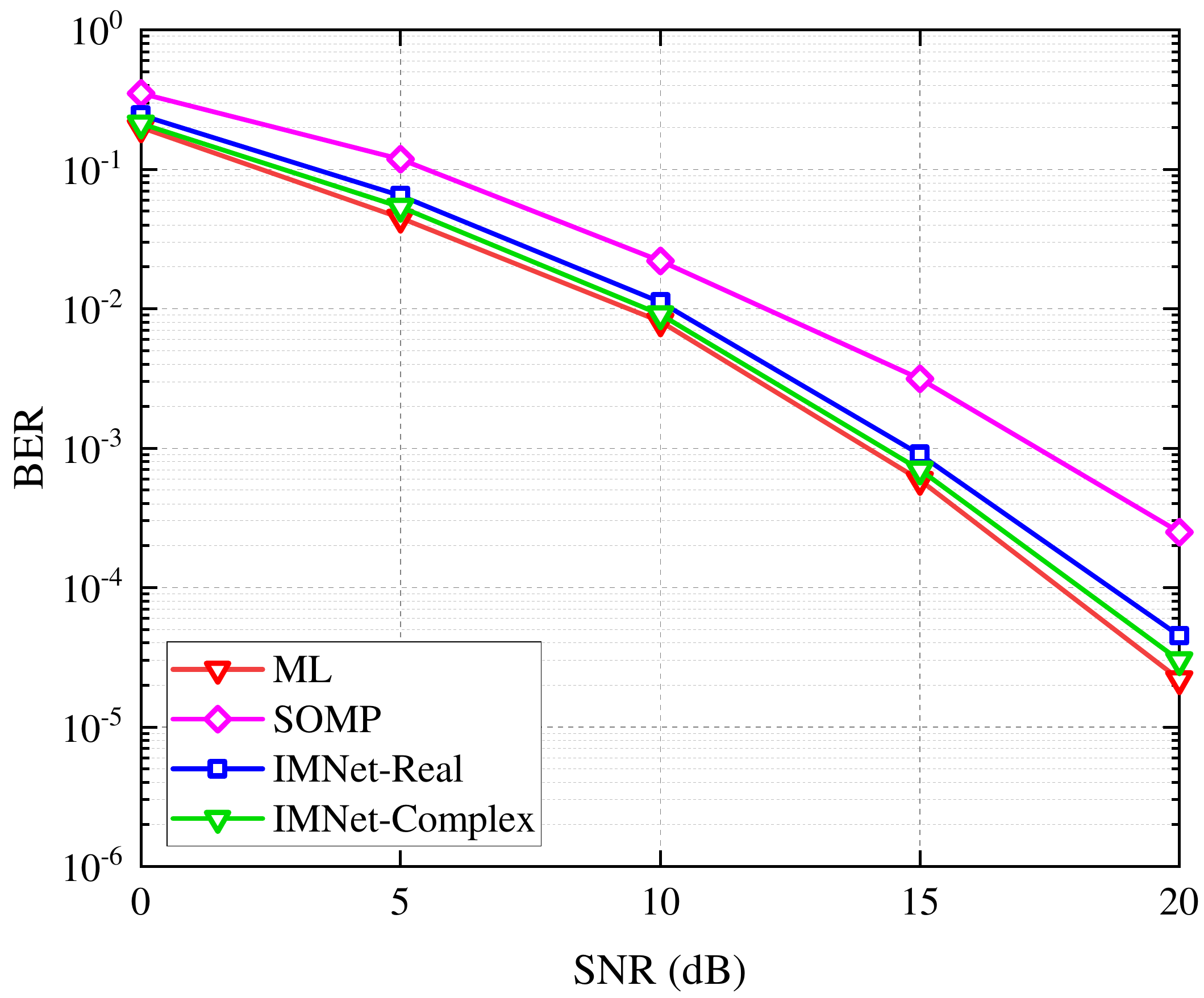}
	}
	\caption{BER performance under Rayleigh fading MIMO channel with perfect channel estiamtion ($N_t = 4, N_u = 1$).}
	\captionsetup{name={Fig.},labelsep=period,singlelinecheck=off,font={small}}
	\label{BER-Ray}
\end{figure*}

\subsection{Simulation Results}\label{Results}

\emph{1) Performance under Rayleigh fading MIMO channel with perfect  CSI:} Fig. \ref{BER-Ray} shows the BER performance of the proposed IMRecoNet and two reference algorithms under Rayleigh fading MIMO channel with perfect channel estimation. The TA configuration is $N_t = 4, N_u = 1$. From the figure, we can find that the ML based detector has the optimal performance both in  4QAM and 16QAM, because it traverses all possible combinations of the legal TACs and the actually transmitted signal vectors. It is obvious that whether it is real-valued IMRecoNet or complex-valued IMRecoNet, its BER performance is significantly ahead of the SOMP signal detection algorithm. And the proposed IMRecoNet has the performance close to the optimal ML based detector. Besides, the complex-valued IMRecoNet has a better performance than the real-valued IMRecoNet with the same network structure. In the perfect CSI case, the BER performance of complex-valued IMRecoNet has an average of 0.77dB gain than the real-valued IMRecoNet. When it comes to 16QAM, the BER performance gain increases into 1.31dB.  

\begin{figure}[htbp]
	\centering
	\includegraphics[width=0.4\textwidth]{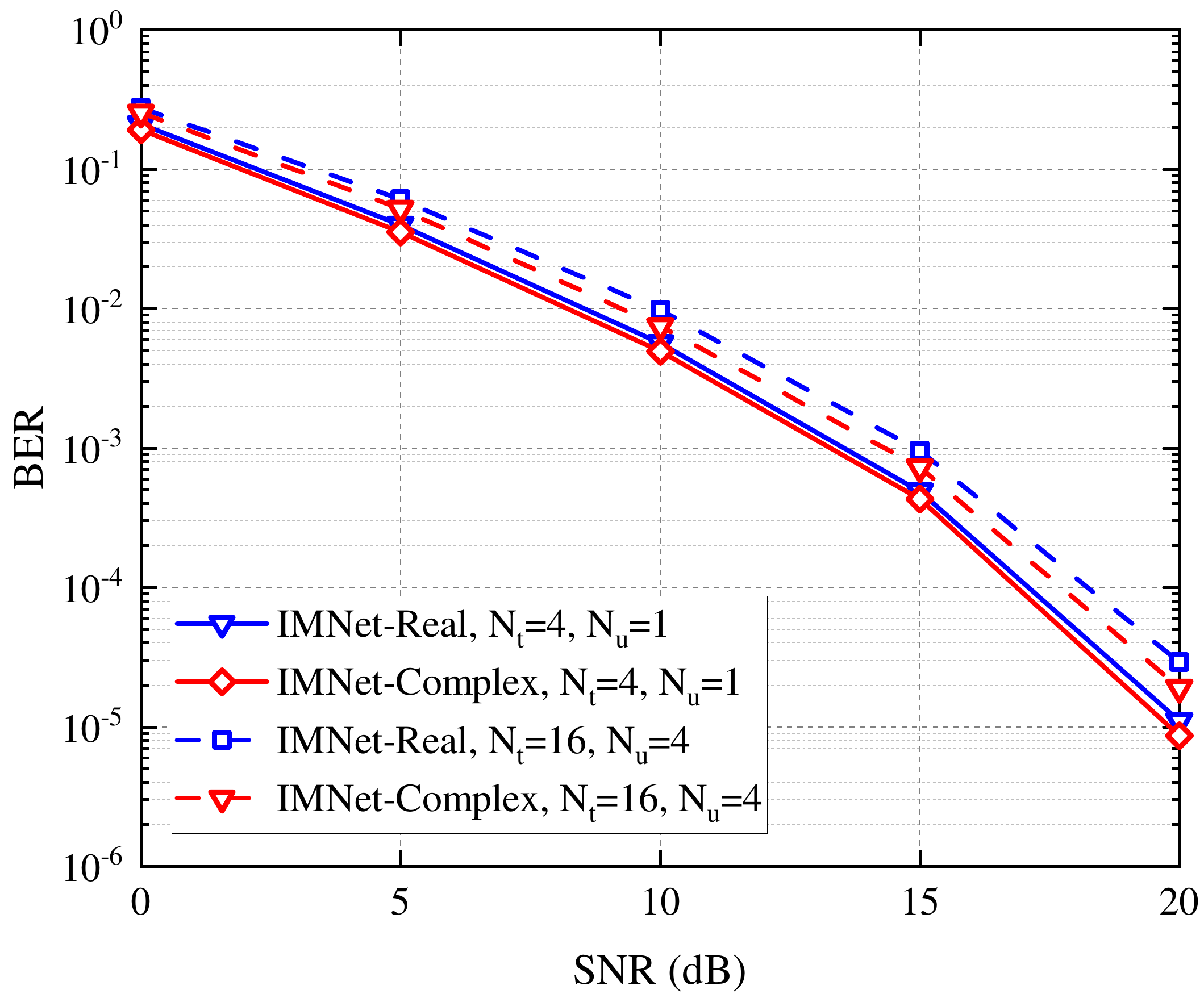}
	  \captionsetup{name={Fig.},labelsep=period,singlelinecheck=off,font={small}}
	\caption{The BER performance of the proposed IMRecoNet under various TA configurations.}
	\label{fig:IMNet_Cases_BER}
\end{figure}

\emph{2) Performance under various TA configurations:} Fig. \ref{fig:IMNet_Cases_BER} shows the BER performance of the proposed IMRecoNet under various TA configurations in Rayleigh fading MIMO channel. The modulation type is 4QAM. It can be seen from Fig. \ref{fig:IMNet_Cases_BER} that IMRecoNet can achieve good BER performance in the two TA configurations, i.e., $N_t = 4, N_u = 1$ and $N_t = 16, N_u = 4$, which indicates that IMRecoNet has good robustness to changes in the number of TAs. On the whole, when the TA configuration changes from $N_t = 4, N_u = 1$ to $N_t = 16, N_u = 4$, the BER performance of IMRecoNet has a certain drop. This is because that when $N_t = 4, N_u = 1$, only one TA is active for data transmission, which means no interference. When $N_t = 16, N_u = 4$, there are four TAs transmitting data at the same time, which has interference between different TAs. The BER performance of the complex-valued IMRecoNet is better than that of the real-valued IMRecoNet in both two cases. When $N_t = 4, N_u = 1$, the BER performance of the complex-valued IMRecoNet has an average gain of 0.70dB than the real-valued IMRecoNet. When the TA configuration becomes $N_t = 16, N_u = 4$, the performance gain is expanded to 1.30dB.

\begin{figure*}[htbp]
	\centering
	\subfigure[]{\includegraphics[width=0.38\textwidth]{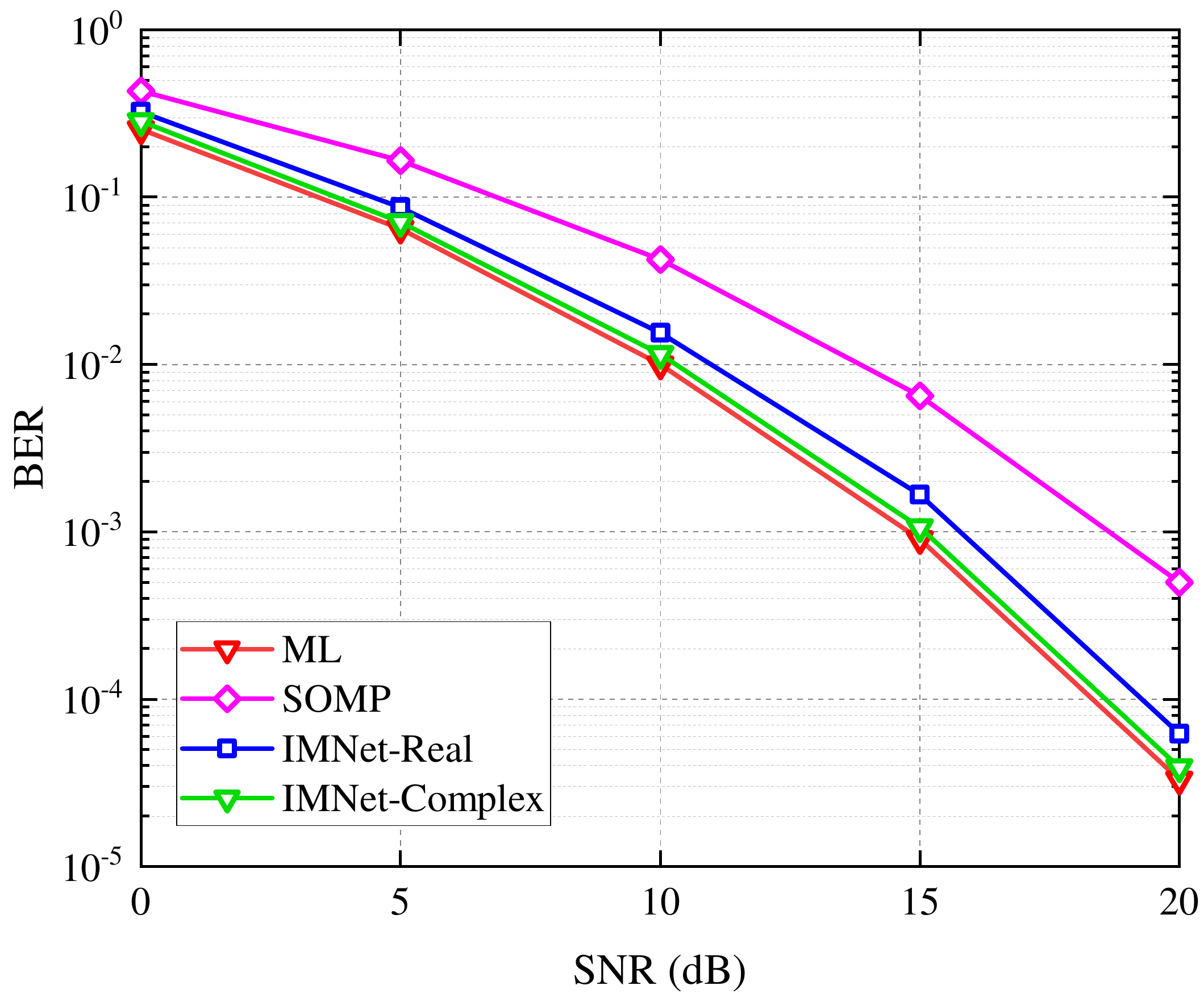}\label{fig:IMNet_Corr_BER}}
	\subfigure[]{\includegraphics[width=0.38\textwidth]{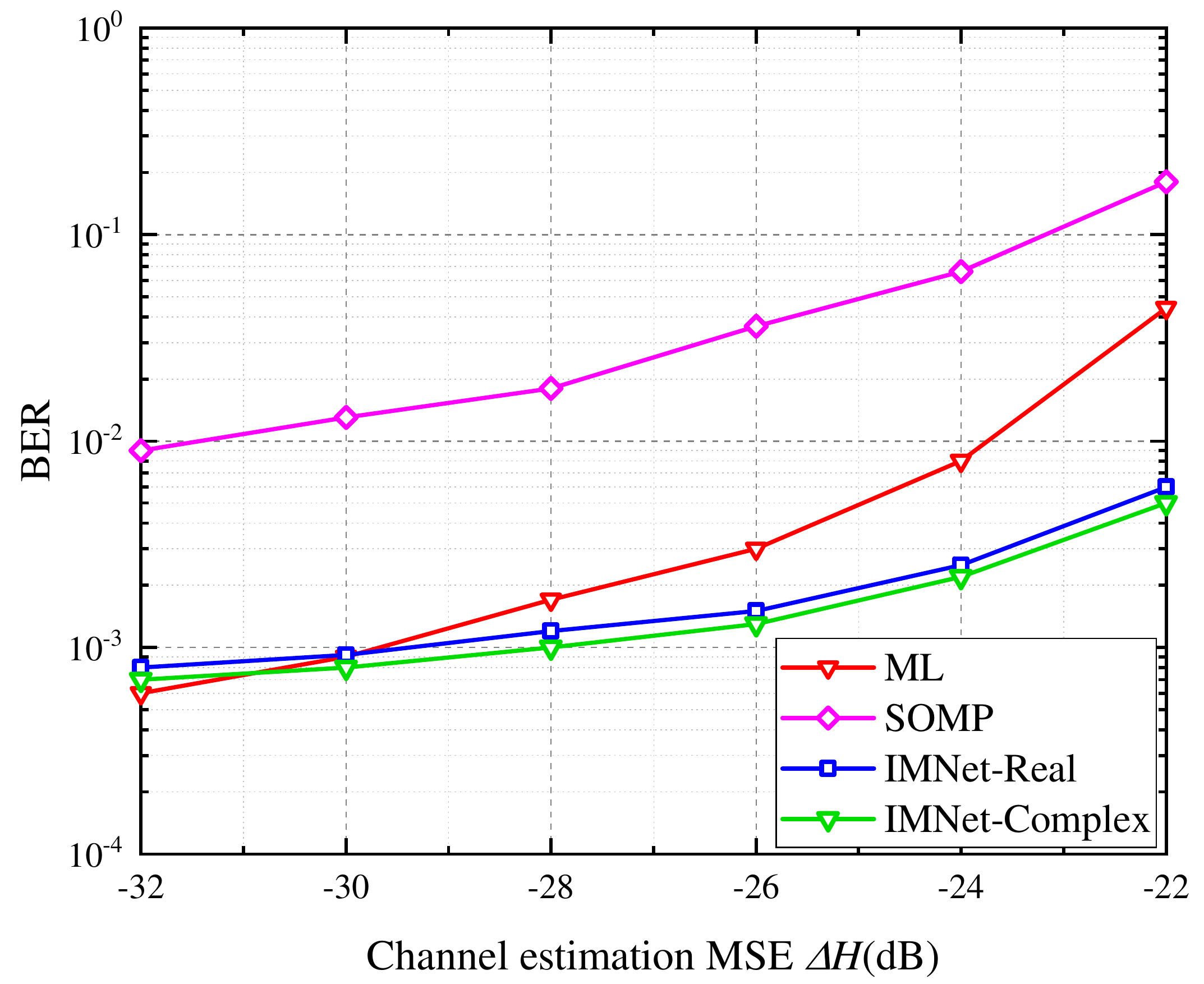}\label{fig:IMNet_Error_BER}}	
	  \captionsetup{name={Fig.},labelsep=period,singlelinecheck=off,font={small}}
	\caption{(a).BER performance under correlated MIMO channel with perfect channel estimation ($N_t = 16, N_u = 4$). (b).BER performance with channel estimation.}	
\end{figure*}

\emph{3) Performance under correlated MIMO channel with perfect channel estimation:} To verify the robustness of the proposed IMRecoNet, we evaluate its performance under correlated MIMO channel modeled by Eq. (\ref{corrMIMOFormulation}) with perfect channel estimation. TA configuration is set to $N_t = 16, N_u = 4$, the correlation cofficient $\rho$ of $\mathbf{R}_r$ and $\mathbf{R}_t$ are both set to 0.5, and the modulation type is 16QAM. From Fig. \ref{fig:IMNet_Corr_BER}, it is clear that the proposed IMRecoNet still maintain the BER performance  close to the optimal ML based detection algorithm. Beside, the BER performance of the proposed IMRecoNet under correlated MIMO channel is still significantly ahead of the BER performance of the SOMP based signal detection algorithm. The results fully show that the proposed IMRecoNet has good robustness to correlated MIMO channel.

\emph{4) Robustness to the channel estimation:} In order to evaluate the robustness of IMRecoNet to channel estimation errors, we consider the following imperfect channel estimation: $\hat{\mathbf{H}}=\mathbf{H}+\Delta\mathbf{H}$, where $\Delta\mathbf{H}$ is the channel estimation error and $\Delta\mathbf{H}\sim\mathcal{CN}(0,\sigma_c^2)$. We use $\hat{\mathbf{H}}$ instead of $\mathbf{H}$ as the input of IMRecoNet for signal detection. Fig. \ref{fig:IMNet_Error_BER} shows the BER performance under 4QAM for different $\sigma_c$. In this simulation, TA configuration is $N_t = 4, N_u = 1$, and SNR is set to 15dB. It can be seen from the result that, as the channel estimation error increases, the BER performance of each algorithm decreases. However, the BER performance of IMRecoNet is far superior to the SOMP based signal detection algorithm, and is very close to the performance of the ML based signal detection algorithm. When channel estimation error is less than -30dB, the BER performance of IMRecoNet is slightly worse than that of the optimal signal detection algorithm based. When the channel estimate is greater than -30dB, the BER performance of IMRecoNet will begin to outperform the performance of the ML based signal detection algorithm. 
\textcolor{blue}{This is becuase the mapping relationship between the received signal and the antenna activation mode is directly learned in deep learning based IMRecoNet instead of obtaining  channel state information via channel estimation.} 
This result shows that IMRecoNet has a certain robustness to channel estimation errors, and also shows that SE Subnet can reduce signal detection error caused by calculation error.

%
%
%

\begin{table}[]
	\centering
		\caption{Comparison of the parameter size and storage size between real-valued IMRecoNet and complex-valued IMRecoNet}
		\label{tab:IMRecoNet_Parameters}
	\begin{tabular}{c|cccl}
		\cline{1-4}
		& \multicolumn{3}{c}{IMNet-Real}                                           &  \\ \cline{1-4}
		& \multicolumn{1}{c|}{AD subnet} & \multicolumn{1}{c|}{SE subnet} & Total  &  \\ \cline{1-4}
		Parameter size & \multicolumn{1}{c|}{1.337M}    & \multicolumn{1}{c|}{0.036M}    & 1.373M &  \\
		Storage size    & \multicolumn{1}{c|}{5.099MB}    & \multicolumn{1}{c|}{0.139MB}    & 5.238MB &  \\ 
		\cline{1-4}
		& \multicolumn{3}{c}{IMNet-Complex}                                           &  \\ \cline{1-4}
		& \multicolumn{1}{c|}{AD subnet} & \multicolumn{1}{c|}{SE subnet} & Total  &  \\ \cline{1-4}
		Parameter size & \multicolumn{1}{c|}{0.668M}    & \multicolumn{1}{c|}{0.019M}    & 0.687M &  \\
		Storage size    & \multicolumn{1}{c|}{2.549MB}    & \multicolumn{1}{c|}{0.073MB}    & 2.622MB & \\	
		\cline{1-4}
	\end{tabular}
\end{table}

\begin{figure*}[htbp]
	\centering
	\subfigure[perfect CSI]{
		\includegraphics[width=0.4\linewidth]{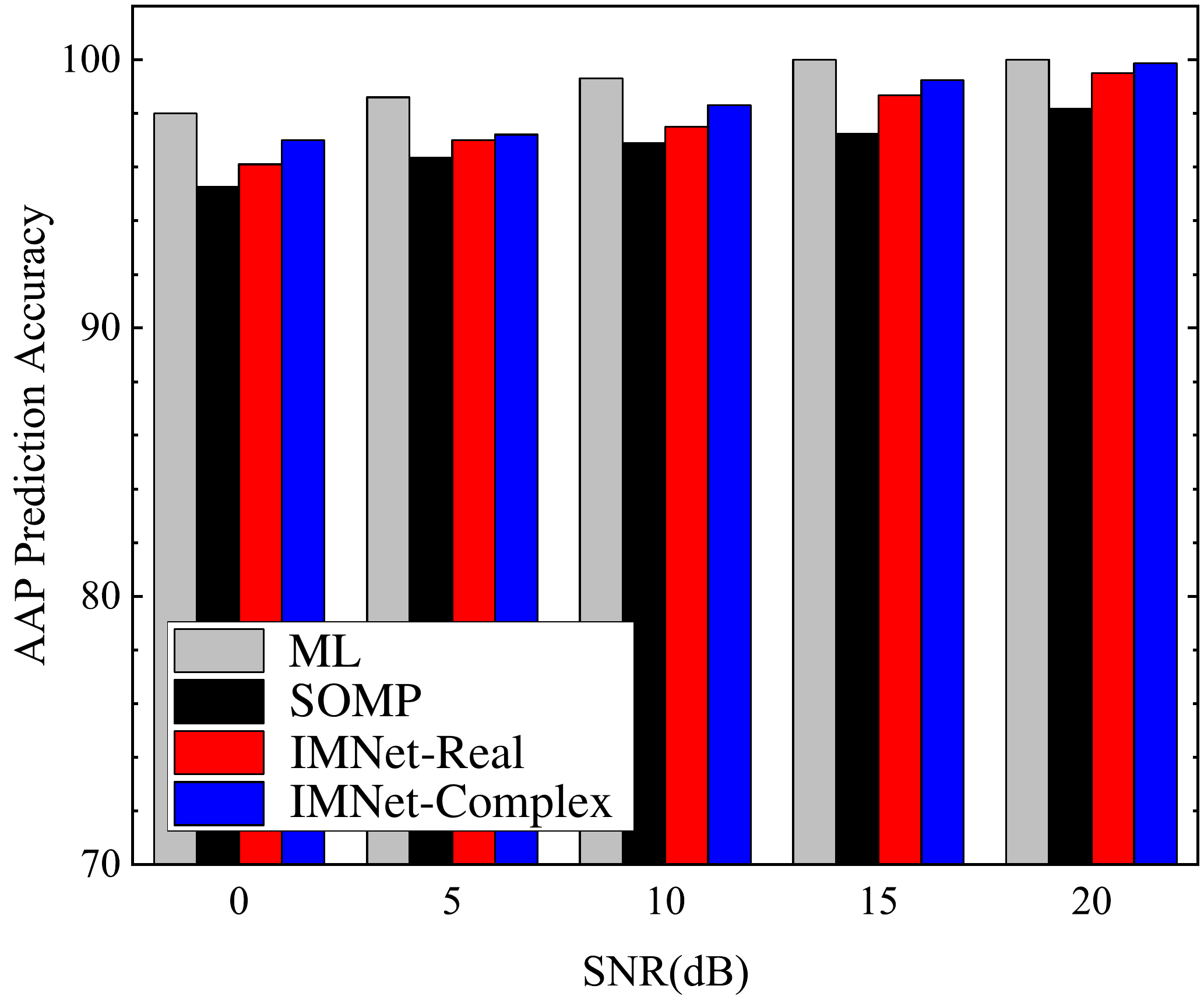}\label{fig:AAP_perfect}
	}
	\subfigure[imperfect CSI]{
		\includegraphics[width=0.4\linewidth]{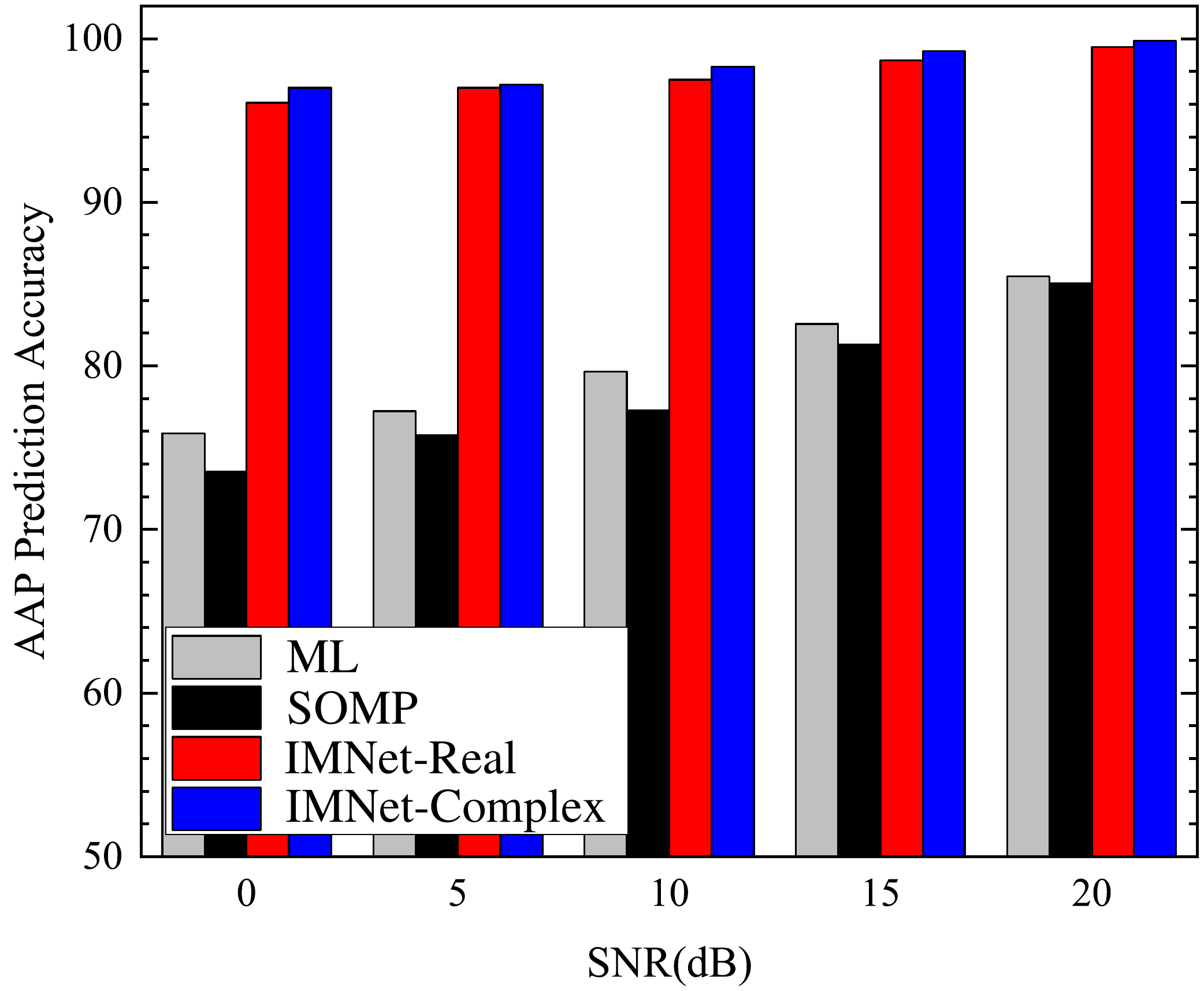}\label{fig:AAP_imperfect}
	}
	\captionsetup{name={Fig.},labelsep=period,singlelinecheck=off,font={small}}
	\caption{\textcolor{blue}{AAP prediction accuracy under Rayleigh fading MIMO channel with perfect/imferfect CSI ($N_t = 4, N_u = 1$).}}
	\label{fig:CR}
\end{figure*}
\begin{figure*}[htbp]
	\centering
	\subfigure[AAPD Subnet]{
		\includegraphics[width=0.4\linewidth]{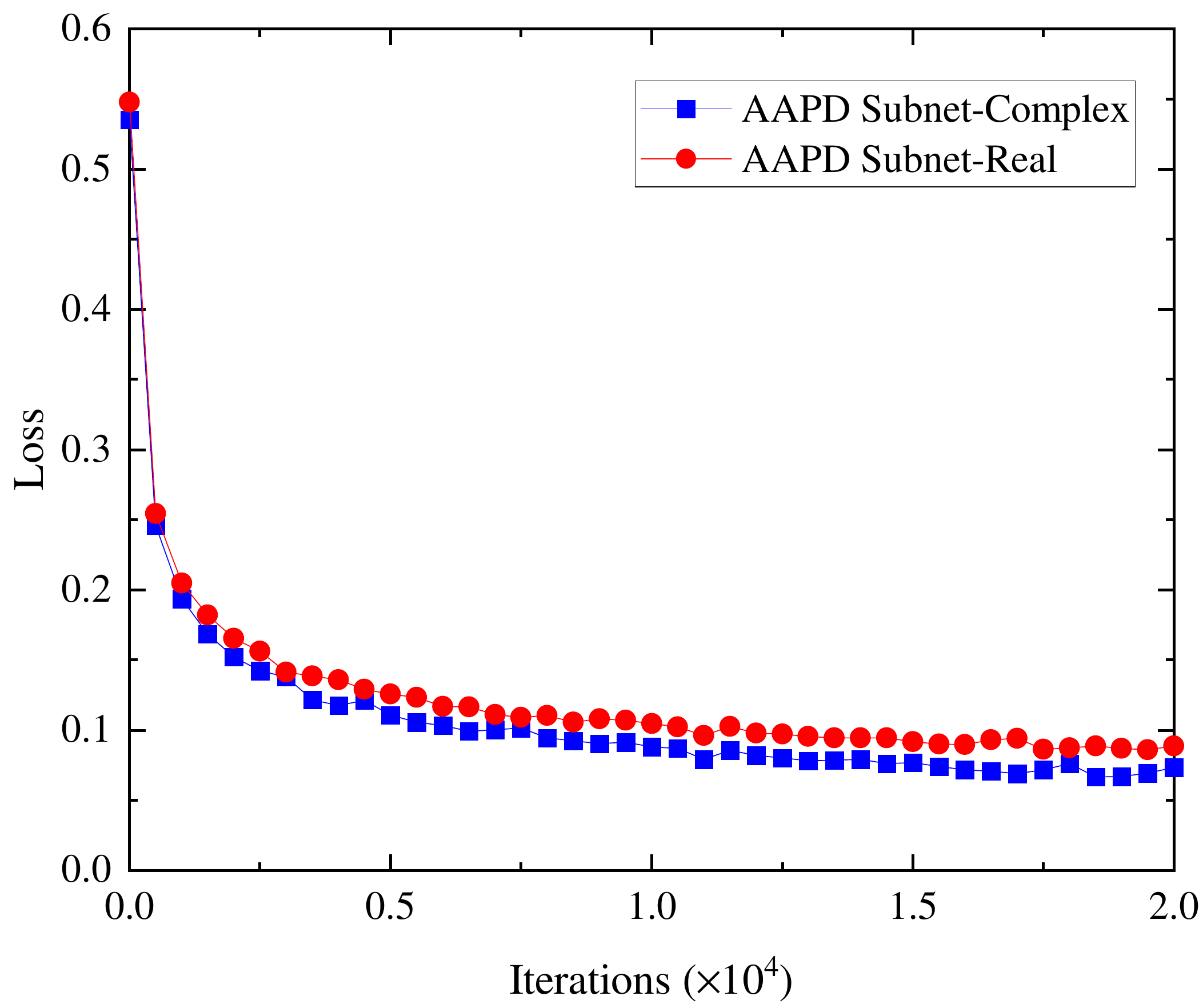}\label{fig:AAPD_Loss}
	}
	\subfigure[SE Subnet]{
		\includegraphics[width=0.4\linewidth]{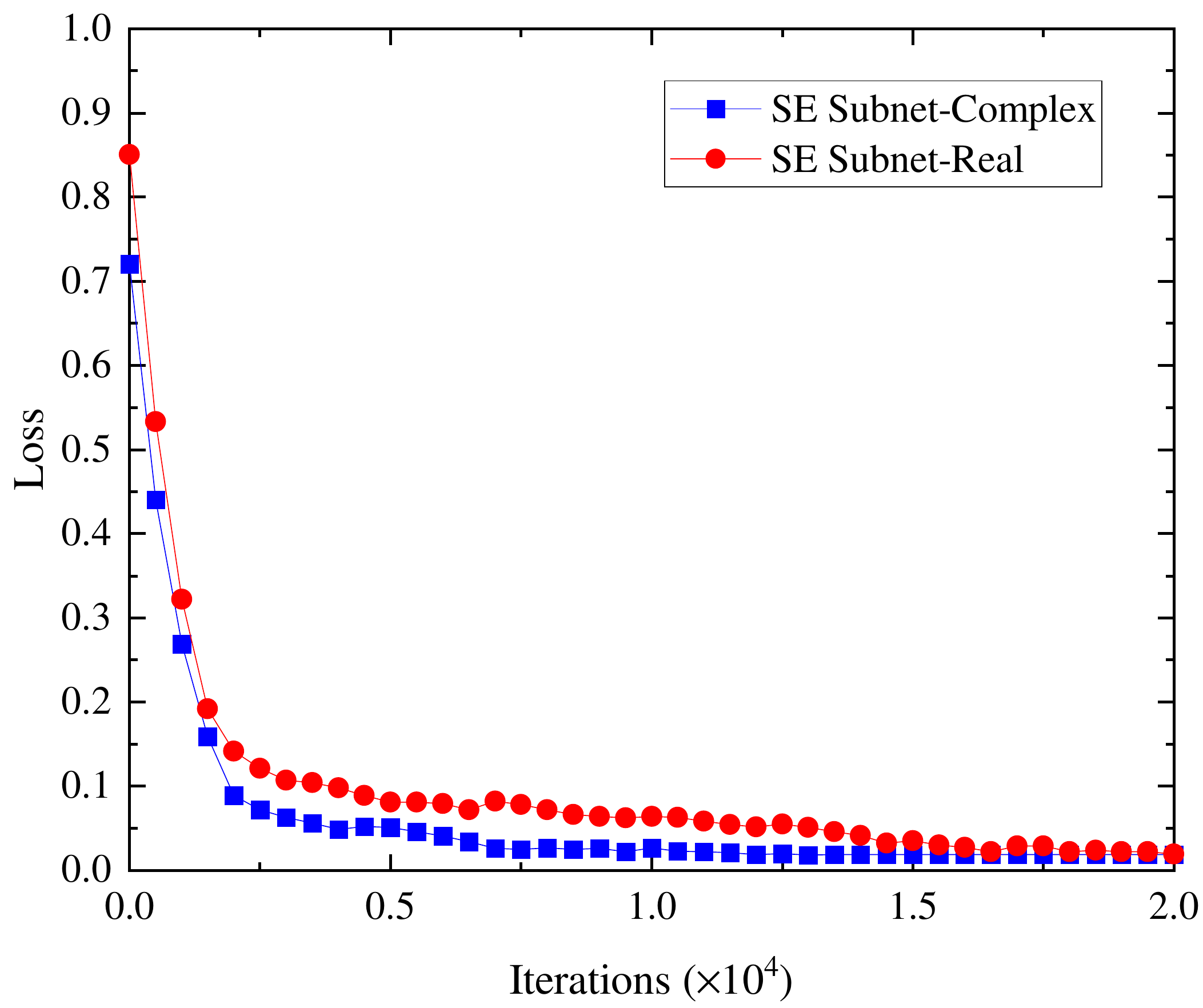}\label{fig:SE_Loss}
	}
	  \captionsetup{name={Fig.},labelsep=period,singlelinecheck=off,font={small}}
	\caption{Convergence speed of real-valued IMRecoNet and complex-valued IMRecoNet.}
	\label{fig:CR}
\end{figure*}
\textcolor{blue}{\emph{5) AAP Prediction Accurancy comparison under perfect CSI and imperfect CSI:}
Fig.\ref{fig:AAP_perfect} and Fig.\ref{fig:AAP_imperfect} present the AAP prediction accuracy of the proposed IMRecoNet and two
baseline algorithms under Rayleigh fading MIMO channel with perfect CSI and imperfect CSI,respectively. TA configuration is $N_t = 4, N_u = 1$ and the modulation mode is 16QAM. We can find that the optimal ML based detector and SOMP detector have about $10\%,20\%$ performance degradation. Compared Fig.\ref{fig:AAP_imperfect} with Fig.\ref{fig:AAP_perfect}, the IMRecoNet has little performance degradation under imperfect CSI. The reason is that ML and SOMP based detector need the CSI as input while our proposed IMRecoNet does not need CSI. Besides, it can be seen that IMNet-Complex performs better than the IMNet-Real due to the complex-valued operations.}

\emph{6) Performance comparison for real-valued IMRecoNet and complex-valued IMRecoNet:} Table \ref{tab:IMRecoNet_Parameters} shows the parameter size and the stoarge size of real-valued IMRecoNet and complex-valued IMRecoNet. The TA configuration is $N_t = 4, N_u = 1$. It can be found from the table that the parameter size of complex-valued IMRecoNet is basically half of that of real-valued IMRecoNet, which is consistent with the previous theoretical analysis. Thanks to the greatly reduced parameter size, the storage size of complex-valued IMRecoNet is only half of the complex-valued IMRecoNet. The introduction of complex-valued operations greatly reduces the parameter size and the storage size of the neural network model, which further increases the possibility of the deployment and use of the neural network in the actual wireless communication system.

The parameter size determines the complexity of the model. In addition to occupying less storage space, a smaller parameter size also has a certain impact on the convergence speed of the network during training period. Fig. \ref{fig:CR} illustrates the convergence speed, which shows the  iterations in the AAPD and the SE subnet when the TA is configured with $N_t = 4, N_u = 1$, the modulation mode is 16QAM, and the SNR is set to 5dB. \textcolor{blue}{In each iteration, the parameters of CNN were updated based on gradient descent.}  Fig. \ref{fig:AAPD_Loss} shows that the complex-valued AAPD subnet has the similar convergence speed to the real-valued AAPD subnet. However, the complex-valued AAPD subnet can obtain a lower loss value than the real-valued AAPD subnet when they converge, which also means the complex-valued AAPD subnet will have better performance. It can be found from Fig.\ref{fig:SE_Loss} that, although the loss value of the complex-valued SE subnet and the real-valued SE subnet are basically the same when converging, the complex-valued SE subnet has a faster convergence speed. The complex-valued SE subnet has basically converged after 10,000 iterations, while the real-valued SE subnet reaches the state of convergence after 15,000 iterations.

\begin{figure}[htbp]
	\centering
	\includegraphics[width=0.4\textwidth]{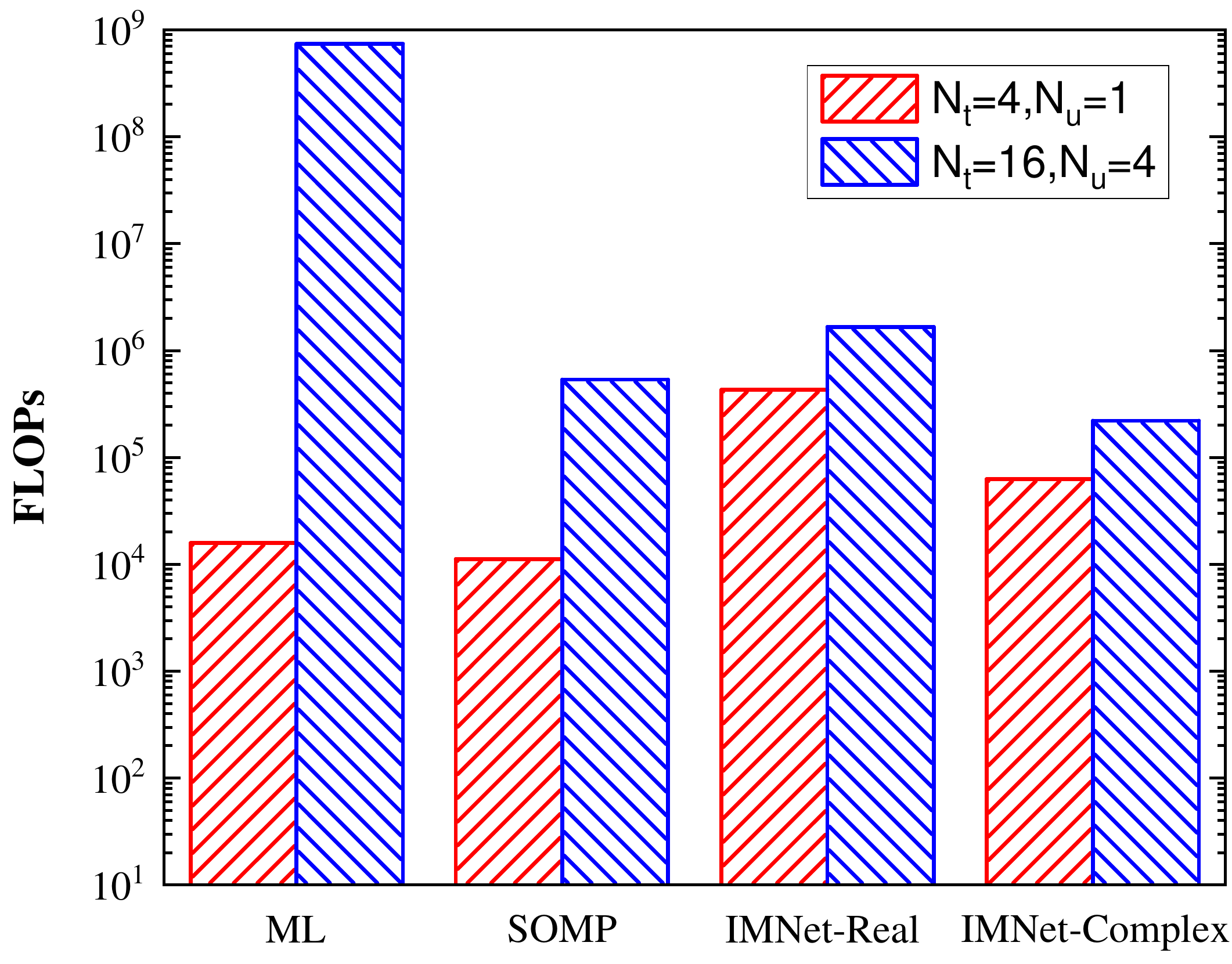}
	\captionsetup{name={Fig.},labelsep=period,singlelinecheck=off,font={small}}
	\caption{FLOPs for each algorithm.}
	\label{fig:IMRecoNet_FLOPs}
\end{figure}

\emph{7) Complexity analysis and comparison:} In order to further compare the complexity of each algorithm, Fig. \ref{fig:IMRecoNet_FLOPs} shows the number of floating point operations (FLOPs) required for signal detection for each algorithm in $N_t = 4, N_u = 1$ and $N_t = 16, N_u = 4$, respectively. It can be seen from the figure that, when the number of TA is small (i.e., $N_t = 4, N_u = 1$), the ML based signal detection algorithm and the SOMP based signal detection algorithm have lower FLOPs. When the number of TAs increase to $N_t = 16, N_u = 4$, the FLOPs of the ML based algorithm increased sharply, much larger than other signal detection algorithms. The FLOPs of SOMP based signal detection algorithms also shows a significant increase. When $N_t = 16, N_u = 4$, the FLOPs of the complex-valued IMRecoNet are lower than the FLOPs of the ML based signal detection algorithm and the SOMP based signal detection algorithm, and also lower than the FLOPs of the real-valued IMRecoNet, which again illustrates that the introduction of the complex-valued operations has played a key role in reducing the complexity of the model.


\section{Conclusion}\label{Conclusion}
In this paper, we design a detector for IM-MIMO systems from a machine learning perspective. Inspired by the huge advances in artificial intelligence in recent years, we propose to integrate deep learning to the design of detector for IM-MIMO systems. Different from assuming the neural network as a black-box and directly using it to detect, we first model the transmission of IM-MIMO systems as a compressed sensing process and formulate the detection process at the receiver as a sparse reconstruction problem. Based on the formulated sparse reconstruction problem, we integrate CNN and a traditional linear MIMO detector (i.e., LS), called IMRecoNet, to realize the detection process. In order to adapt the neural network to complex signal processing in wireless communications, complex value operations (i.e., complex convolution and complex activation) are further introduced to the design of IMRecoNet. Extensive simulations have been carried out to verify the performance of real value and complex value based IMRecoNet under various scenarios. The BER performance shows that IMRecoNet achieves better performance than the traditional detectors, and the introduction of complex value operations can indeed further improve performance.

The results of this paper has shown the great potential of deep learning in the physical layer of wireless communications. Compared with computer vision and natural language processing, wireless communication has its own unique characteristics, such as complex value signals and strict delay requirements. Therefore, it is inefficient and unrealistic to directly introduce deep learning into wireless communications. In this paper, we first introduce complex-valued operations to CNN for adapting the characteristic that signal in wireless communications is represented as complex values. Compared with CNN with real value operations, the introduction of complex operations leads to certain performance improvement. Last but not least, only complex convolution and simple complex activation are implemented in this paper, how to design other deep learning components that match the characteristics of wireless communications is still open and challenging.

\bibliographystyle{IEEEtran}
\bibliography{IMRecoNet}

\end{document}